# The contribution of intermolecular spin interactions to the London dispersion forces between chiral molecules

M. Geyer,[†] R. Gutierrez,[†,a] V. Mujica,[‡,*] J. F. Rivas Silva,[††] A. Dianat,[†] G. Cuniberti[†,¶]

[†] Institute for Materials Science and Max Bergmann Center of Biomaterials, Dresden University of Technology, 01062 Dresden, Germany
[‡] Arizona State University, School of Molecular Sciences, PO Box 871604, Tempe, Arizona 85287-1604, USA
[¶] Dresden Center for Computational Materials Science and Center for Advancing Electronics Dresden, TU Dresden, 01062 Dresden, Germany
[††] Instituto de Física Luis Rivera Terrazas, Benemérita Universidad Autónoma de Puebla, Apdo. Postal J48, Col. San Manuel, Puebla, Pue. C. P. 72570, Mexico
[*] Kimika Fakultatea, Euskal Herriko Unibertsitatea and Donostia International Physics Center (DIPC), P. K. 1072,20080 Donostia, Euskadi, Spain.
[a] Corresponding autor: *rafael.gutierrez@tu-dresden.de*

**ABSTRACT**

Dispersion interactions are one of the components of van der Waals (vdW) forces, which play a key role in the understanding of intermolecular interactions in many physical, chemical and biological processes. The theory of dispersion forces was developed by London in the early years of quantum mechanics. However, it was only in the 1960s that it was recognized that for molecules lacking an inversion center such as chiral and helical molecules, there are chirality-sensitive corrections to the dispersion forces proportional to the rotatory power known from the theory of circular dichroism and with the same distance scaling law $R^{-6}$ as the London energy. The discovery of the Chirality-Induced Spin Selectivity (CISS) effect in recent years has led to an additional twist in the study of chiral molecular systems, showing a close relation between spin and molecular geometry. Motivated by it, we propose in this investigation to describe the mutual induction of charge and spin-density fluctuations in a pair A-B of chiral molecules by a simple physical model. The model assumes that the same fluctuating electric fields responsible for vdW forces can induce a magnetic response via a Rashba-like term, so that an spin-orbit field acting on molecule B is generated by the electric field arising from charge density fluctuations in molecule A (and viceversa). Within a second-order perturbative approach, these contributions manifest as an effective intermolecular exchange interaction. Although expected to be weaker than the standard London forces, these interactions display the same $R^{-6}$ distance scaling.

## I. INTRODUCTION

Dispersion forces, together with Keesom and Debye forces, are the three pillars building the van-der-Waals (vdW) interactions, which are ubiquitous in physics, chemistry and biology, and of fundamental importance for understanding weak intermolecular interactions.[1,2] While Keesom and Debye forces can be explained within classical physics, only the advent of the quantum theory allowed to introduce and rationalize dispersion forces. In fact, the quantum mechanical theory of dispersion interactions was developed by London[3] and by London and Eisenchitz[4] already in the early years of quantum mechanics. Their approach was based on a second-order perturbation treatment describing the interaction between fluctuating dipole moments in a pair of atoms at distance $R$, and it naturally led to an attractive contribution to the potential energy with an $R^{-6}$ distance dependence. Meanwhile, very sophisticated methodological developments have taken place to account for vdW interactions in the modeling of physical systems. This includes pairwise additive approaches, treatments of the correlation energy based on the adiabatic-connection fluctuation-dissipation (ACFD) approach, many-body corrections to account for non-additivity, and long-range pairwise nonlocal functionals, among others, see e.g. Refs.[5,6,7] for reviews.

Interestingly, slightly earlier than the work by London and Eisenchitz, Rosenfeld[8] formulated a theoretical framework to describe natural optical activity. The Rosenfeld equation provided the basis for quantifying experiments in optical circular dichroism, by relating the difference in absorption of right- and left circularly polarized light by chiral molecules to the imaginary part of a product of transition matrix elements of the electric dipole and the magnetic dipole operators: $\text{Im}\{\langle 0|\mu_{el}|n\rangle\langle n|\mu_{mag}|0\rangle\}$.[9]

A natural question which may arise is whether the interaction between chiral molecules -in particular dispersive interactions- may also allow for chiral discrimination. Theoretical research in this direction started as early as in the 1960s by Craig, Power, and Thirunamachandran,[10] and has been continued mostly in the works by Woolley,[11] Barron,[9,12] and Salam.[13,14] Methodologically, these investigations use either second-order perturbation theory (as in London's paper) or so–called molecular quantum electrodynamics (mQED). A fundamental result of these studies is that there exists a chirality-dependent contribution to dispersion forces, involving a susceptibility tensor related to the rotatory power obtained in the Rosenfeld theory of circular dichroism.[8]

Also closely associated to chiral molecules is the recently discovered Chirality-Induced Spin Selectivity (CISS) effect,[15,16,17,18,19,20,21,22,23,24,25,26,27] which suggests a non-trivial connection between spin and geometry in chiral molecules. Theoretically, a large number of studies based on model Hamiltonian[28,29,30,31,32,33,34,35,36,37,38,39,40,41,42,43,44,45,46] and, more recently, first-principle[47,48,49,50] approaches have been presented. A common starting point in the vast majority of the presented models so far is the presence of a specific spin-orbit coupling contribution connected with the helical geometry (either postulated or obtained through a coarse-graining procedure). Still, the ultimate origin of the CISS effect remains a matter of intense debate, see e.g. the recent discussions in Refs.[39,51,52] From a general point of view, the CISS effect implies the presence of correlated charge and spin density fluctuations in a chiral molecule under the influence of an external perturbation, which can be related to a propagating charge as in electron transport experiments, to electron transfer, to the proximity of magnetized substrates or, as we will address here, to the presence of another chiral molecule, where fluctuating fields can induce both an electric and a magnetic response mediated by spin-orbit interactions.

A hint that CISS-related exchange interactions may provide a contribution to the interaction energy *between* chiral molecules was presented in the combined experimental-theoretical study by Kumar et al.[21], where a Density-Functional theory (DFT) calculation revealed a chirality-sensitive energy contribution for the interaction between a pair of molecules. The obtained contribution had its origin in short-range electronic exchange effects, which have also been shown, using spin-polarized DFT, to lead to the onset of spontaneous magnetization in arrays of chiral molecules.[48] More recently, San Sebastian et al.[23] demonstrated the existence of an enantiospecific response in Nuclear Magnetic Resonance experiments for a class of optically active metal−organic frameworks. Although no full theoretical description was developed in the paper, it was nevertheless suggested that besides the well-known *J-J* coupling required to describe spin-spin interactions in NMR setups, an additional intermolecular contribution might be needed to account for chirality-induced electron spin polarization.

Motivated by these different perspectives, we may wonder if CISS-related phenomena may still be relevant for dispersion interactions beyond the range where electronic exchange, and thus the antisymmetry of the many-body wave function, plays a role. The problem is clearly highly non-trivial. From the perspective of a first-principle approach, dispersion interactions are included in the formally exact expression of the electronic correlation energy.[53] The latter can be written e.g., using the ACFD approach, in terms of charge density response functions, the bare Coulomb interaction, and a non-local exchange-correlation (XC) kernel $f_{XC}$.[53] Hence, any possible corrections to dispersion forces mediated by CISS-related fluctuations could in principle be accounted for, if spin-dependent response functions and the spin-dependent XC-kernel could be exactly computed (in a non-collinear framework) or if appropriate approximations could be found for them. We know, however, that this is not the case even for the London interactions (where the XC-kernel is usually neglected in the spirit of the RPA approximation), so that a large number of approaches with different levels of sophistication has been developed to properly account for dispersion in practical DFT calculations.[5,6,7,54,55,56,57] Similarly, CISS transport physics has only been partially captured by the effective mean-field, single-particle DFT calculations performed so far,[47, 49,50] although short-range exchange effects between chiral molecules can still be described with conventional XC-functionals.[48] However, in the vdW range it is not obvious how to scrutinize the influence the CISS-effect could have on dispersion.

We do not attempt to tackle this many-body problem in its full complexity. Instead, we suggest a simple phenomenological way of connecting charge fluctuations in one chiral molecule with a spin-dependent response in the other. We combine the basic idea behind the origin of dispersion -charge fluctuation generating to lowest order a dipole field- with the general rationale behind CISS -correlated charge and spin-density fluctuations- to assume that the fluctuating electric field of one of the molecules can act as the source of a spin-orbit field in the second molecule and vice versa, thus influencing the magnetic response of the latter. We have called this proposed contribution *induced intermolecular spin-orbit (SO) interaction*. In the frame of mQED approaches, this intermolecular SO could be seen as an additional correction to the magnetic density fluctuation term, which only contributes for chiral molecular systems due to the underlying symmetries, specifically the breaking of space inversion symmetry.[10,58]

Interatomic and intermolecular spin-orbit effects have been discussed some time ago in the context of the exciton problem in molecular crystals,[59,60] and in intersystem crossing processes during the collision of argon atoms with small organic molecules.[61] Meath and Hirschfelder also discussed, on the basis of a multipole expansion of the Breit-Pauli Hamiltonian, the distance scaling of intermolecular forces, including relativistic corrections.[62] Their formalism was, however, very general, and it is not obvious how to apply it in the context of the CISS effect in chiral molecules. Therefore, a simpler model of these intermolecular spin-dependent effects may be appealing. More recently, the influence of intra- and inter-nuclear spin-orbit coupling has been addressed in the study of excited states of heteronuclear alkali metal dimers using quasirelativistic electronic wave functions,[63] as well as in the study of the molecular crystal $Mo_3S_7(dmit)_3$ using 4-components relativistic DFT.[64] Although these latter studies are less relevant for our model approach in terms of the addressed systems, we will use the computed values of the intermolecular spin-orbit energies in Ref. [65] as a reference point for the estimation of the coupling constant appearing in our model.

Within our proposed phenomenological, single particle approach, we find a correction to the London dispersion energy arising from the intermolecular SO, which displays, at the level of the dipole approximation, an $R^{-6}$ scaling with the intermolecular separation. The intermolecular SO interaction can be mapped, within our perturbative approach, onto an effective anisotropic pseudo-spin Hamiltonian with an exchange coupling tensor encoding dipole fluctuations. Moreover, in the limit of a pair of linear molecules, these corrections disappear due to the vanishing of the associated response tensors, so that only the standard London dispersion will remain. Despite the phenomenological character of our approach, we believe that our results point to the potential relevance of chirality-related pseudo-spin interactions between helical molecules within the van der Waals interaction range. Our results can therefore be of interest e.g. in addressing phenomenologically, outside the range of exchange-mediated effects, the chiral discrimination component of intermolecular interactions in biological systems, where chiral structures are ubiquitous, and their role in molecular recognition. In this field, full spin polarized DFT calculations are, as a rule, not feasible due to the large number of atoms involved.

In the next section we will introduce the model Hamiltonian for the intermolecular SO (Sec. II.A) and analyze it in second order degenerate perturbation theory (Sec. II.B). In Sec. II.C, a simple continuum helix model is introduced in order to calculate the relevant electrical transition dipole moments. The main results are then presented in Sec. III and a formal generalization of the approach is summarized in Sec. IV. A summary and an outlook of open issues is finally provided in Sec. V.

## II. INDUCED INTERMOLECULAR SPIN-ORBIT COUPLING IN CHIRAL MOLECULES

### A. The model Hamiltonian

We consider two helical molecules $A$ and $B$ (not necessarily of the same kind) described by some general Hamilton operator $H_0^{A,B}$, whose specific structure is not of interest at this stage. To simplify the calculations, we will assume, however, that $H_0^A$ and $H_0^B$ are single-particle Hamiltonians diagonal in spin space, i.e. do not include spin-dependent interactions, an assumption that can be lifted in a more complete calculation. Within the vdW range with intermolecular separations $R \ll \lambdabar$, with $\lambdabar$ being the reduced wavelength of a typical electronic transition, relativistic retardation effects can be neglected and, as a result, static interaction potentials can be used.

As discussed in the Introduction, we aim at proposing a minimal model to account for the interrelated charge- and spin-density fluctuations setting in when two chiral molecules approach each other. This corresponds to a broader view of the CISS effect and its role in intermolecular interactions, beyond transport or transfer processes.[66,20] These dynamical charge density rearrangements are assumed to have an influence on the dispersion forces outside the region of overlapping electronic densities. However, extracting CISS-related contributions within an atomistic framework, e.g. in DFT, goes well beyond our scope. Instead, we attempt at mimicking this rather intricate effect by considering the following model for a pair of helical molecules: charge density fluctuations in one molecule, say $A$, generate a fluctuating electric field, which acts as the source of a spin-orbit field in molecule $B$ (and vice versa). In this way, a phenomenological way of connecting charge and spin effects may be achieved. Although in the electrostatic potential different multipoles can be included, we limit ourselves to treat the problem at the level of the dipole approximation to keep the problem as simple as possible. Then, the proposed model Hamiltonian is given by:

$$H = H_0^A + H_0^B + H_{\text{int}}, \quad (1)$$

$$H_{\text{int}} = H_{d-d} + H_{SO} = -\frac{1}{4\pi\varepsilon_0} \mu_\alpha^A T_{\alpha\beta} \mu_\beta^B + (1+P_{AB})\lambda_{SO} \vec{E}_A(R) \cdot \left( \vec{\sigma}_B \times \vec{p}(B) \right),$$

$$E_{j=A,B,\alpha}(R) = -\frac{1}{4\pi\varepsilon_0} \frac{\mu_\beta(j)}{R^3} (\delta_{\alpha\beta} - 3 r_\alpha r_\beta) = -\frac{1}{4\pi\varepsilon_0} T_{\alpha\beta} \mu_\beta(j).$$

The interaction term $H_{int}$ in Eq. (1) includes the standard dipole-dipole electrostatic interaction $H_{d-d}$ leading to the London dispersion energy and the spin-dependent term $H_{SO}$. The operator $P_{AB}$ exchanges the molecular indices.

In Eq. (1) we have used the Einstein sum convention for Greek indices, which will be kept all along this paper. $\vec{E}_{A(B)}(R)$ is the electric field (within the dipole approximation) of molecule $A(B)$ acting on molecule $B(A)$, $\mu_\beta(j)$ is the $\beta$ component of the corresponding dipole moment operator of molecule $j=A,B$, $r_\alpha = R_\alpha / R$, and $R = |R_A - R_B|$ is the distance between molecules $A$ and $B$. The distance is measured perpendicular to the helical axes, which are assumed to be parallel to each other. The coupling strength $\lambda_{SO}$ is in general different from the bare relativistic coupling $e\hbar/(2mc)^2$,[40,38] and may encode in a non-trivial way intramolecular effects.[40,38]

At this stage, we will introduce atomic units to express the energy scale in units of $\hbar^2/m_e a_B^2$ and the length scales in Bohr radii $a_B$. As a result, we can introduce the dimensionless coupling constant ($\lambda_{SO}$ has dimensions of C s kg$^{-1}$): $\lambda_{eff} = \hbar \lambda_{SO} / e a_B^2 = 2,34 \times 10^5 \times \lambda_{SO}$. Using the Levy-Civita tensor, the interaction part in Eq. (1) can then be rewritten as:

$$H_{int} = H_{d-d} + H_{SO} = -\mu_\alpha^A T_{\alpha\beta} \mu_\beta^B - \lambda_{eff} \varepsilon_{\alpha\beta\gamma}(1+P_{AB})\sigma_\alpha(B)p_\beta(B)T_{\gamma\rho}\mu_\rho(A). \qquad (2)$$

Notice that no symmetrization of the SO-part of the Hamiltonian is required to make it Hermitian, since the momentum operator acts on different degrees of freedom as the dipole operator. We do not assume the molecules to possess a permanent dipole moment, which would lead to additional corrections arising from induction forces. Such contributions can be treated separately. The model in Eq. (1) describes, thus, how a fluctuating charge density (characterized by a dipole operator at this level) on one molecule induces not only a fluctuating charge density on the other molecule (being at the origin of the conventional London forces), but also a spin-orbit field. A realistic estimate of $\lambda_{eff}$ is hard to achieve using first-principle calculations, since the proposed model is largely phenomenological and it is not obvious, as previously mentioned, how to extract the CISS-related response from the correlation energy. Therefore, we leave this parameter open and provide later on in Sec. III.A a rough estimate of its plausible orders of magnitude.

### B. Degenerate perturbation theory

As mentioned in the previous section, the reference Hamiltonians for both molecules will be assumed to be diagonal in spin space, i.e. the zero-order wave functions of the system originate from a (helical) molecule without intrinsic spin-orbit interactions, i.e. spin-orbit coupling in a molecule arises only through the influence of the other molecule. This is a reasonable, though simplified, starting point, since we are not addressing here issues related to spin transport and the resulting spin polarization in helical systems. In a later section we will adopt a specific model for the helix, but for the time being the problem is formulated in a general way.

In the absence of the spin degrees of freedom, a standard second order perturbative approach would be sufficient as commonly used to derive the London dispersion energy. However, in our case each molecular energy eigenvalue in zero order is twofold degenerate because of spin. This implies that for the two interacting molecules there is a fourfold spin degeneracy for each energy eigenvalue $E_{nm} = E_n(A) + E_m(B)$. Within each of these subspaces we can define a tensor product basis of the form $|\uparrow\uparrow\rangle, |\uparrow\downarrow\rangle, |\downarrow\uparrow\rangle, |\downarrow\downarrow\rangle$. The first spin component refers to molecule A, the second one to molecule B. We note, however, that our molecules are "featureless", i.e. no atomistic electronic structure is considered. Hence, these spin states can be considered as a formal way of labelling the degeneracy of the corresponding molecular Hamiltonians, but are not *per se* individual electronic spins belonging to an atom. In this sense, we may more accurately speak of pseudo-spins, but we will use indistinctly both notations in this paper. Only within a more advanced atomistic framework, where spin-orbitals are introduced, can we really associate these variables with electronic spins. This approach is similar in nature to what is done in the Heisenberg model of spin Hamiltonians.

Now, it turns out that perturbative corrections which are first order in the spin-orbit coupling strength from Eq. (1) identically vanish, since the matrix elements of the momentum operator over the ground state are zero (assuming that the molecules do not possess any permanent dipole moments), i.e.: $\langle 0_A\uparrow, 0_B\uparrow | H_{int} | 0_A\uparrow, 0_B\uparrow \rangle \sim \varepsilon_{\alpha\beta\gamma} \langle \uparrow | \sigma_\alpha(B) | \uparrow \rangle \langle 0_B | p_\beta(B) | 0_B \rangle T_{\gamma\rho} \langle 0_A | \mu_\rho(A) | 0_A \rangle = 0$. A similar argument holds for the dipole-dipole part of the interaction Hamiltonian.

As a result, we need to use second-order degenerate perturbation theory.[67] Here, the secular matrix for the perturbation $H_{int}$ needs to be built within each fourfold degenerate spin subspace of the Hilbert space of the interacting molecules, $H = H_A \otimes H_B$, using the previously introduced spinor basis. The obtained eigenvalues of the 4x4 secular matrix will provide the required energy corrections to the non-interacting ground state of the two molecules to second order in the intermolecular spin-orbit coupling strength.

A typical matrix element will, therefore, look like:

$$W^{(2)}_{sp,s'p'}(R) = -\sum_{\substack{n,m\neq 0 \\ r,r'=\uparrow,\downarrow}} \frac{\langle 0_{A,s} 0_{B,p} | H_{int} | n_{A,r} m_{B,r'} \rangle \langle n_{A,r} m_{B,r'} | H_{int} | 0_{A,s'} 0_{B,p'} \rangle}{\omega_{n0} + \omega_{m0}} . \qquad (3)$$

We have used the simplified notation $|n_{A,s} m_{B,p}\rangle = |n_{A,s}\rangle \otimes |m_{B,p}\rangle \equiv |n_A\rangle |\chi_s^A\rangle \otimes |m_B\rangle |\chi_p^B\rangle$, where $|\chi^{j=A,B}\rangle = (\uparrow \quad \downarrow)^T$ are the corresponding spinors. Moreover, $\omega_{m0} = E_m(B) - E_0(B)$ and $\omega_{n0} = E_n(A) - E_0(A)$.

After a lengthy but straightforward calculation, we obtain the following tensor representation of the second-order energy correction (we only show here contributions involving $H_{SO}$, the pure dipole-dipole correction related to the London dispersion is spin-diagonal and standard, and is presented in the Supplementary Material):

$$W^{(2)}(R) = \delta_{\alpha\beta} 1_A \otimes 1_B (Q^{BA}_{\alpha\beta} + Q^{AB}_{\alpha\beta}) \qquad (4)$$
$$+ \left\{ 1_A \otimes \mathrm{Im}(\sigma^B_\xi h^{BA}_\xi) + \mathrm{Im}(\sigma^A_\xi h^{AB}_\xi) \otimes 1_B \right\}$$
$$+ \left\{ 1_A \otimes i Q^{BA}_{\alpha\beta} \varepsilon_{\alpha\beta\xi} \sigma^B_\xi + i Q^{AB}_{\alpha\beta} \varepsilon_{\alpha\beta\xi} \sigma^A_\xi \otimes 1_B \right\}$$
$$+ \sigma^A_\alpha \otimes \sigma^B_\beta (K^{BA}_{\alpha\beta} + K^{AB}_{\alpha\beta}).$$

In Eq. (4) we have introduced the tensors $Q^{BA}_{\alpha\beta}$ and $K^{BA}_{\alpha\beta}$ and the complex vector $h^{BA}_\xi$. These quantities include different response functions, which are defined in the following way:

$$Q^{BA}_{\alpha\beta} = -\lambda^2_{eff} \varepsilon_{\alpha\kappa\gamma} \varepsilon_{\beta\kappa'\gamma'} T_{\gamma\rho} T_{\gamma'\rho'} \sum_{n,m\neq 0} \frac{1}{\omega_{n0} + \omega_{m0}} p^{0m}_\kappa(B) \mu^{0n}_\rho(A) p^{m0}_{\kappa'}(B) \mu^{n0}_{\rho'}(A) = \qquad (5)$$

$$= -\lambda^2_{eff} \varepsilon_{\alpha\kappa\gamma} \varepsilon_{\beta\kappa'\gamma'} T_{\gamma\rho} T_{\gamma'\rho'} \sum_{n,m\neq 0} \frac{\omega^2_{m0}}{\omega_{n0} + \omega_{m0}} \mu^{0m}_\kappa(B) \mu^{m0}_{\kappa'}(B) \mu^{0n}_\rho(A) \mu^{n0}_{\rho'}(A) =$$

$$= -\lambda^2_{eff} \varepsilon_{\alpha\kappa\gamma} \varepsilon_{\beta\kappa'\gamma'} T_{\gamma\rho} T_{\gamma'\rho'} \int_0^\infty \frac{du}{2\pi} \underbrace{\left\{ 2 \sum_{m\neq 0} \frac{\omega^3_{m0}}{u^2 + \omega^2_{m0}} \mu^{0m}_\kappa(B) \mu^{m0}_{\kappa'}(B) \right\}}_{\Gamma^B_{\kappa\kappa'}(iu)} \underbrace{\left\{ 2 \sum_{n\neq 0} \frac{\omega_{n0}}{u^2 + \omega^2_{m0}} \mu^{0n}_\rho(A) \mu^{n0}_{\rho'}(A) \right\}}_{\alpha^A_{\rho\rho'}(iu)}$$

$$\equiv -\lambda^2_{eff} \varepsilon_{\alpha\kappa\gamma} \varepsilon_{\beta\kappa'\gamma'} T_{\gamma\rho} T_{\gamma'\rho'} \int_0^\infty \frac{du}{2\pi} \Gamma^B_{\kappa\kappa'}(iu) \alpha^A_{\rho\rho'}(iu).$$

The second row in Eq. (5) is obtained by using the general relation (in atomic units): $\langle n | p_\alpha | m \rangle = i(E_n - E_m)\langle n | r_\alpha | m \rangle = -i\omega_{mn}\langle n | \mu_\alpha | m \rangle$, to express the momentum matrix elements in terms of dipole matrix elements. In the third row, we have used the identity:

$$\frac{1}{a+b} = \frac{2}{\pi} \int_0^\infty du \frac{ab}{(a^2 + u^2)(b^2 + u^2)}.$$

to define the two susceptibility tensors of the last row of Eq. (5).

In a similar way the *K*-tensor in Eq. (4) is given by:

$$K^{BA}_{\alpha\beta} = -\lambda^2_{eff} \varepsilon_{\alpha\kappa\gamma} \varepsilon_{\beta\kappa'\gamma'} T_{\gamma\rho} T_{\gamma'\rho'} \sum_{n,m\neq 0} \frac{1}{\omega_{n0} + \omega_{m0}} p^{0n}_\kappa(A) \mu^{0m}_\rho(B) p^{m0}_{\kappa'}(B) \mu^{n0}_{\rho'}(A) = \qquad (6)$$

$$= -\lambda^2_{eff} \varepsilon_{\alpha\kappa\gamma} \varepsilon_{\beta\kappa'\gamma'} T_{\gamma\rho} T_{\gamma'\rho'} \sum_{n,m\neq 0} \frac{\omega_{n0}\omega_{m0}}{\omega_{n0} + \omega_{m0}} \mu^{0n}_\kappa(A) \mu^{0m}_\rho(B) \mu^{m0}_{\kappa'}(B) \mu^{n0}_{\rho'}(A) =$$

$$= -\lambda^2_{eff} \varepsilon_{\alpha\kappa\gamma} \varepsilon_{\beta\kappa'\gamma'} T_{\gamma\rho} T_{\gamma'\rho'} \frac{1}{2\pi} \int_0^\infty du \underbrace{\left\{ 2 \sum_{m\neq 0} \frac{\omega^2_{m0}}{u^2 + \omega^2_{m0}} \mu^{0m}_\rho(B) \mu^{m0}_{\kappa'}(B) \right\}}_{\Pi^B_{\rho\kappa'}(iu)} \underbrace{\left\{ 2 \sum_{n\neq 0} \frac{\omega^2_{n0}}{u^2 + \omega^2_{m0}} \mu^{0n}_\kappa(A) \mu^{n0}_{\rho'}(A) \right\}}_{\Pi^A_{\kappa\rho'}(iu)}$$

$$\equiv -\lambda^2_{eff} \varepsilon_{\alpha\kappa\gamma} \varepsilon_{\beta\kappa'\gamma'} T_{\gamma\rho} T_{\gamma'\rho'} \int_0^\infty \frac{du}{2\pi} \Pi^B_{\rho\kappa'}(iu) \Pi^A_{\kappa\rho'}(iu).$$

Finally,

$$h_\xi^{AB} = -\lambda_{eff}\varepsilon_{\xi\eta\kappa}T_{\alpha\beta}T_{\kappa\eta}\int_0^\infty \frac{du}{2\pi}\underbrace{\left\{2\sum_{m\neq 0}\frac{\omega_{m0}^2}{u^2+\omega_{m0}^2}\mu_\alpha^{0m}(B)\mu_\eta^{m0}(B)\right\}}_{\Pi_{\alpha\eta}^B(iu)}\underbrace{\left\{2\sum_{n\neq 0}\frac{\omega_{n0}}{u^2+\omega_{m0}^2}\mu_\beta^{0n}(A)\mu_\rho^{n0}(A)\right\}}_{\alpha_{\beta\rho}^A(iu)} \quad (7)$$

$$\equiv -\lambda_{eff}\varepsilon_{\xi\eta\kappa}T_{\alpha\beta}T_{\kappa\eta}\int_0^\infty \frac{du}{2\pi}\Pi_{\alpha\eta}^B(iu)\alpha_{\beta\rho}^A(iu).$$

The response tensors $\Pi$, $\Gamma$, and $\alpha$ are defined at imaginary frequencies $\omega=iu$ via analytic continuation into the upper half-plane. Going back now to Eq. (4), the first term is diagonal in spin space. The second line in Eq. (4) arises from mixed contributions involving matrix elements of $H_{d-d}$ and $H_{SO}$. Notice that this term is linear in $\lambda_{eff}$. The third line in Eq. (4) can be rewritten as an effective magnetic Zeeman-like interaction acting separately on the A- and B-subspaces:

$$i\varepsilon_{\alpha\beta\xi}1_A\otimes\sigma_\xi^B Q_{\alpha\beta}^{BA} = i1_A\otimes\sigma_\xi^B(\varepsilon_{\alpha\beta\xi}Q_{\alpha\beta}^{BA}) = i1_A\otimes\sigma_\xi^B B_\xi = i1_A\otimes\boldsymbol{\sigma}^B\cdot\boldsymbol{b}^{BA}. \quad (8)$$

Lastly, the fourth row has the form of a generalized effective exchange interaction with coupling given by $J_{\alpha\beta} = K_{\alpha\beta}^{BA} + K_{\alpha\beta}^{AB}$. Then, we obtain the more compact:

$$W^{(2)}(R) = 1_A\otimes 1_B Tr(Q^{BA}+Q^{AB}) + \left\{1_A\otimes \text{Im}(\boldsymbol{\sigma}^B\cdot\boldsymbol{h}^{BA}) + \text{Im}(\boldsymbol{\sigma}^A\cdot\boldsymbol{h}^{AB})\otimes 1_B\right\} \quad (9)$$
$$+ \left[1_A\otimes\boldsymbol{\sigma}^B\cdot i\boldsymbol{b}^{BA} + \boldsymbol{\sigma}^A\cdot i\boldsymbol{b}^{AB}\otimes 1_B\right] + J_{\alpha\beta}\sigma_\alpha^A\otimes\sigma_\beta^B,$$

which is the central result of our investigation. Notice that all terms are time-reversal invariant.

Due to the symmetry property $Q^{BA} = (Q^{AB})^\dagger$ and $K^{BA} = (K^{AB})^\dagger$ the $Q$- and $J$-tensors are both Hermitian. However, if the zero-order wave functions are real, then it turns out that $\boldsymbol{b}^{BA} = \boldsymbol{b}^{AB} \equiv 0$, since they involve contributions with the following form: $\varepsilon_{\alpha\beta\xi}Q_{\alpha\beta}^{BA} + \varepsilon_{\beta\alpha\xi}Q_{\beta\alpha}^{BA} = \varepsilon_{\alpha\beta\xi}Q_{\alpha\beta}^{BA} - \varepsilon_{\alpha\beta\xi}Q_{\beta\alpha}^{BA} = (\varepsilon_{\alpha\beta\xi} - \varepsilon_{\alpha\beta\xi})Q_{\alpha\beta}^{BA} = 0$. Similarly $\text{Im}\,\boldsymbol{h}^{BA} \equiv 0$, because the vector $\boldsymbol{h}^{BA}$ will have real components.

On the other hand, if the zero-order Hamiltonian has complex eigenstates (e.g. when considering intra-molecular spin-orbit coupling), then $Q$- and $J$-tensors and the vector $\boldsymbol{h}$ are not real-valued and all terms in Eq. (9) need to be considered. In this study we will only address the case of real-valued wave functions and, hence, the energy correction to the dispersion interaction adopts the simpler form:

$$W^{(2)}(R) = (1_A\otimes 1_B)Tr(Q^{BA}+Q^{AB}) + J_{\alpha\beta}\sigma_\alpha^A\otimes\sigma_\beta^B. \quad (10)$$

We have thus mapped the proposed phenomenological intermolecular SO interaction from Eq. (1) onto an effective anisotropic pseudo-spin exchange contribution plus a spin-diagonal contribution. Notice that all interaction tensors in the general expression given by Eq. (9) as well as in Eq. (10) display an $R^{-6}$ scaling with the intermolecular distance due to the scaling of the product $\boldsymbol{T}\otimes\boldsymbol{T}$ in Eqs. (5), (6) and (7). In the Supplementary Material we show, by exploiting the Unsöld approximation,[68] that the $K$-tensor and, hence, the $J$-tensor can be written as $-C_{\alpha\beta}^{BA}R^{-6}$, where the coefficient is roughly proportional to the product of ground state matrix elements of the electric dipole fluctuations of each molecule $\sim \langle 0|\mu^2(A)|0\rangle\langle 0|\mu^2(B)|0\rangle$. This better highlights the connection between charge density fluctuations and spin-spin (or pseudo-spin) interactions in Eq. (10). We stress that both Eq. (9) and Eq. (10) represent the second-order energy correction as a tensor in the (pseudo)-spin Hilbert space of the combined system. This tensor will become a 4x4 matrix with the current choice of spin basis.

### C. Continuum model of a helix

The results obtained so far did not assume any specific underlying model for the zero-order Hamiltonian. Moreover, there is no information on the helical symmetry in the Hamiltonian of Eq. (1). To further proceed and obtain explicit numerical results for the various response tensors previously defined as well as a dependence on the helical geometry, we have to formulate a microscopic model for the reference molecular systems.

As mentioned in Sec. II.A, the assumed zero order Hamiltonian is a single-particle operator diagonal in spin space. We will base our further analysis on the model introduced by Tinoco[69] with Hamilton operator given by: $H = -(\hbar^2/2m)\partial^2/\partial s^2$, where $s$ is the arc length along the helix. For a helix of finite length this model is equivalent to a particle in a box model and it clearly yields the same type of eigenfunctions, which read, using open boundary conditions: $\Psi_n(s) = \sqrt{2/L}\sin(k_n s)$. Here, $L$ is the length of a helix with radius $R_0$ and pitch $b$, and given by $L = KL_0 = K\sqrt{b^2 + 4\pi^2 R_0^2}$. $K$ is the number of helical turns and the discrete wave vector $k_n = \pi n/L$. Using these wave functions, the transition dipole matrix elements for a helix whose axis is parallel to the z-axis in a local coordinate system can be computed as:

$$\mu_1^{nm} = 0, \tag{11}$$

$$\mu_2^{nm} = \frac{8}{\pi} s_h k_n R_0 \frac{mn[(-1)^{m+n}-1]}{[(m-n)^2 - 4K^2][(m+n)^2 - 4K^2]},$$

$$\mu_3^{nm} = \frac{4}{\pi^2} k_n |b| \frac{mn[(-1)^{m+n}-1]}{(m^2-n^2)^2},$$

$$\mu^{nm} = (\pm R_0/2, 0, 0) \quad \text{for } 2K = m \mp n, m \geq n,$$

with $s_h = \pm 1$ denoting the helix chirality. We obtain then for products of the transition dipole matrix elements (notice that the ground state "0" corresponds in this model to the $n=1$ quantum number):

$$\mu_\alpha^{1m}\mu_\beta^{m1} = 16[1+(-1)^m]^2(Km)^2 \times$$

$$\times \begin{pmatrix} 0 & 0 & 0 \\ 0 & \dfrac{4R_0^2/\pi^2}{[(m-1)^2 - 4K^2]^2[(m+1)^2 - 4K^2]^2} & \dfrac{2s_h |b| R_0/\pi^3}{[(m-1)^2 - 4K^2][(m+1)^2 - 4K^2](m^2-1)^2} \\ 0 & \dfrac{2s_h |b| R_0/\pi^3}{[(m-1)^2 - 4K^2][(m+1)^2 - 4K^2](m^2-1)^2} & \dfrac{b^2/\pi^4}{(m^2-1)^2} \end{pmatrix}_{\alpha\beta}, \tag{12}$$

for $m \neq 2K+1$ and $m \neq 2K-1$. For the specific case $m = 2K \pm 1$, we obtain:

$$\mu_\alpha^{1m} \mu_\beta^{m1} = \frac{R_0^2}{4} \begin{pmatrix} 1 & 0 & 0 \\ 0 & 0 & 0 \\ 0 & 0 & 0 \end{pmatrix}_{\alpha\beta}.$$

We mention that this purely 1D model can be extended to include transversal degrees of freedom due to a finite size confinement potential. However, as long as no transitions between the transversal states are considered (space adiabaticity), the presented results remain valid.[10] We stress that this model is the simplest one that can catch basic features of the helical symmetry, but it is clear that it cannot describe intra-molecular many-body or intramolecular spin-orbit effects. Therefore, it should be considered as a first step towards a more advanced treatment of this problem.

### D. Dispersion potential

For the numerical evaluation we will use a simple Lennard-Jones type of potential to include the induced spin-orbit related corrections to the dispersion interactions:

$$V_{LJ}(s_h^A, s_h^B, R) = \frac{C_0}{R^{12}} - \frac{C_6 + C_{SO}(s_h^A, s_h^B, R)}{R^6} = \frac{C_0}{R^{12}} - \frac{C_{6,eff}}{R^6}. \qquad (13)$$

The constant $C_0$ is arbitrary and only controls the strongly repulsive part of the potential, which is not described our approach. The coefficient $C_6 = C_{London}^{isotr}$ relates to the standard London dispersion interaction. We have computed this coefficient using the same helix model in Sec. 1 of the Supplementary Material. The label *isotr* means that we perform an isotropic approximation for the London contribution, so that it does not depend on the chirality in order to better see the influence of the last term, the (pseudo)spin- and helicity-dependent correction $C_{SO}(s_h^A, s_h^B, R)$, which includes the influence of the intermolecular SO and does depend on the molecular chirality $s_h$. It is discussed and calculated in the next section. Due to the dipole approximation, this second contribution represents basically a correction to the $C_6$ coefficient, leading to a new $C_{6,eff}$. We stress at this point that we are interested in spatial regions where vdW interactions are dominant and eventually lead to local minima, which can be qualitatively described by a Lennard-Jones potential. More elaborate implementations could use e.g. damping functions for the short-range part of the potential, but this is not relevant to the present level of description. We also remark that the $C_6$ coefficient is not related here to a specific atomic species, but it rather describes the polarizability of the underlying helix model. Therefore, a direct comparison with experimental results or computed $C_6$ coefficients for specific atom types is not possible.

## III. RESULTS

### A. Parameter estimation

The effective coupling strength $\lambda_{eff}$ does not need to be identical with the bare relativistic spin-orbit parameter ($\lambda_{rel}=5,64\times 10^{-11}$ C s kg$^{-1}$). However, an accurate estimation is difficult without implementing the necessary corrections at a first principle level. Therefore, to provide an orientation about the orders of magnitude we are dealing with, we have carried out Density-Functional Theory (DFT) calculations for a homo- and a heterochiral molecular pair (all DFT results and technical details are presented in Sec. 2 of the Supplementary Material). The main goal of these DFT calculations is to stress the fact that dispersion terms must be included to obtain any chiral-discrimination effect, which points out to the intricate nature of the connection between exchange and spin-dependent interactions as those involved in the CISS effect.

We have chosen helicene as a simple carbon-based helical molecular system, which does not have a sizeable permanent dipole moment in the ground state, so that the main contribution to vdW interactions is expected to arise from dispersion interactions. We have considered two situations: (a) a D-D helicene pair and (b) a D-L pair. Here, we focus on the most relevant *Case 3* (as presented in Sec. 2 of the Supplementary Material), where the contribution of *intramolecular* spin-orbit effects on the dispersion energy was estimated (see Tables **S3-S6**, and in particular Table **S7** in the Supplementary Material). For the situations (a) and (b) the helical axis were enforced to be parallel to each other and we did not perform any optimization of the relative orientation or the intermolecular separation (in contrast to *Case 1* in Sec. 2 of the Supplementary Material). We computed then the dispersion energy contribution in (a) and (b) with and w/o intramolecular spin-orbit. It is important to note that for the dispersion corrections we used the Tkatchenko-Scheffler (TS) approach, where the dispersion coefficients and damping functions are dependent on the charge density,[70] and will hence be indirectly affected by switching on and off spin-orbit interaction terms. This is in contrast with the Grimme parametrizations, which only account for changes in the local environment of a given atom due to its insertion e.g. into a molecule.

For a typical separation $R$=3.73 Å between the molecule pair, we obtain for case (a) $|E_{disp}^{int,SO} - E_{disp}^{int,no-SO}|=13$ meV and for case (b) $|E_{disp}^{int,SO} - E_{disp}^{int,no-SO}|=0,09$ meV . This difference does not vary, however, monotonously with the separation due to a delicate interplay between SO and structural relaxation. Still, these results provide an order of magnitude estimate of the influence of spin-orbit coupling on the dispersion energy and show that the effect is not trivial. However, since the treated SO contributions are intramolecular, the obtained energy scales can only be considered, in the best case, as a rough upper bound for the strength of the proposed intermolecular SO interaction, which are clearly expected to be much smaller.

In Ref.[64] a four-component relativistic DFT study on the molecular crystal Mo$_3$S$_7$(dmit)$_3$ was carried out and it was shown, that both intra- and intermolecular spin-orbit interactions were significant. Clearly, we are treating here a different setup, but we may still consider the intermolecular contributions computed in Ref.[64] to provide an additional reference point for the orders of magnitude estimates. As shown in Table **1** of that reference, all intermolecular contributions are on average smaller than 0.5 meV.

Based on these two sets of DFT calculations, we will *assume* much smaller energy scales for the intermolecular SO interaction $\delta E_{SO}$ by a factor $10^3$ -$10^4$. We can then provide an approximate value for $\lambda_{eff}$ . To proceed, we will assume that only transition dipole moments $\mu_{12}$ connecting the ground state with the first excited state are relevant. Then, we can use Eq. (11) to get the corresponding values or alternatively, what we follow now, use computed values for helicene at the DFT level, as shown in Table **1** of Ref.[71] (the difference in values for the two ways of calculation is relatively small and does not affect orders of magnitude estimates as we aim here). From Table **1** in Ref.[71] we can write

$\mu_{12} = \mu_{Hel} \times 10^{-19}$ esu·cm=$(\mu_{Hel}/10)$ Debye, with $\mu_{Hel}$ being a numerical factor of order 1. We then use e.g. Eq. (6) in the isotropic limit (for simplicity) to get $\delta E_{SO} \sim |J| = 2K(R)$, and from here an estimate of $\lambda_{eff} = (\sqrt{3}/2)(R^3/|\mu_{12}|^2)\sqrt{|J|/\omega_{12}}$, where all the quantities are in atomic units. Assuming now a helix with equal pitch and radius $b=R_0 \cong 0,4$ nm= 7,55 a.u., a similar value for the intermolecular separation $R \cong 0,4$ nm, and with $\omega_{12} = 3\pi^2/L^2$, we can get $\lambda_{eff} \sim 30$ when assuming $\delta E_{SO} \sim 2 \times 10^{-4}$ meV. However, we stress that this parameter is in general dependent on the molecular system. We also remark that using the bare relativistic SO interaction would lead to a $\lambda_{eff}$ 4-5 orders of magnitude smaller.

We can also provide an order of magnitude estimate of the ratio $E_{GS}^{(2)}/W_{London}^{(2),isotr}$, where $E_{GS}^{(2),isotr}$ is the state with the lowest energy obtained from the diagonalization of Eq. (10). The label *isotr* means, as previously stated, that we exploit an isotropic approximation for the sake of simplicity (see Sec. 1 of the Supplementary Material) in the tensors from Eqs. (5) and (6). Otherwise, it would be impossible to get analytical estimates. As done in the previous paragraph, we assume that only the lowest excitation energy is contributing to the sums in the corresponding susceptibility tensors (transition $n=1 \rightarrow n=2$, see Eq. (11)). Based on this, we obtain:

$$\eta = \frac{W_{SO}^{(2),isotr}}{W_{London}^{(2),isotr}} = \frac{22}{27}\lambda_{eff}^2 \omega_{12}^2 = \frac{22}{3}\pi^4 \left(\frac{\lambda_{eff}}{K^2 L_0^2}\right)^2 = \frac{22}{3K^4}\pi^4 \left(\frac{\lambda_{eff}}{b^2 + 4\pi^2 R_0^2}\right)^2. \tag{14}$$

Assuming again $b=R_0 \equiv \rho$, one first gets:

$$\eta = 0,435 \times \left(\frac{\lambda_{eff}}{K^2 \rho^2}\right)^2. \tag{15}$$

Taking now $\rho \approx 0,4$ nm=7,55 $a_B$, $K=1$ and $\lambda_{eff} = 20-30$ yields $\eta \sim 0,002-0,006$, which is small but not necessarily negligible.

We finally notice that an even smaller choice of $\lambda_{eff}$ can be achieved by considering the presence of a gap in the eigenvalue spectrum between the ground state ($n=1$) and the first excited state ($n=2$) as a simple way of mimicking a HOMO-LUMO gap in a molecule. In this case, the first excitation energy of the continuum helix Hamiltonian will be given by: $\omega_{12} = \Delta + 3\pi^2/L^2$. Now, instead of Eq. (14) we obtain the following expression:

$$\eta = \frac{22}{27}(\Delta\lambda_{eff})^2 [1 + \frac{3\pi^2}{\Delta K^2 \rho^2 (1+4\pi^2)}]^2 \approx \frac{22}{27}(\Delta\lambda_{eff})^2. \tag{16}$$

As a result, for a typical value of $\Delta = 2$ eV $= 0,073$ a.u., it is enough to have $\lambda_{eff} = 1.5$ to obtain a ratio $\eta \sim 1 \times 10^{-2}$. This in its turn can also lead to an even smaller value $\delta E_{SO}$. This approximation may play a role in more realistic electronic structure calculations for molecular systems, which display a qualitatively different electronic spectrum as that obtained with our simple model and possess HOMO-LUMO gaps in the order of magnitude given above. The previous estimations show that we may obtain non-negligible corrections to dispersion interactions without assuming fully unrealistic energy scales.

We also stress, to close this section, that what will control the relevance of the derived interaction terms in Eq. (10), are charge and spin density *fluctuations*, which are not necessarily small (as reflected in general in the transition matrix elements). Therefore, in a real molecular system the overall effect can be much stronger than what we are discussing based on the simplest possible helix model.

### B. Anisotropic case

If we keep the full anisotropy of the problem, the results are very difficult to interpret due to the strong mixing of the original spinor basis, and a more refined theoretical formulation with a more realistic zero order Hamiltonian may be in place. Since this goes beyond the scope of this investigation, we will introduce some simplifications in Eq. (10) regarding the exchange interaction term as given by the *J*-tensor: (i) for the diagonal components, we assume only two effective coupling parameters $J_{zz}$ an $J_\perp = (J_{xx} + J_{yy})/2$, while (ii) for the off-diagonal coupling we introduce the approximation $J_0 = (J_{xy} + J_{yz} + J_{xz})/3$. Thus, Eq. (10) reduces to the simpler expression:

$$W^{(2),aniso}(R) = 1_A \otimes 1_B Tr(Q^{AB} + Q^{BA}) + J_\perp(\sigma_x^A \otimes \sigma_x^B + \sigma_y^A \otimes \sigma_y^B) + J_{zz}\sigma_z^A \otimes \sigma_z^B \qquad (17)$$
$$+ J_0(\sigma_x^A \otimes \sigma_y^B + \sigma_x^A \otimes \sigma_z^B + \sigma_y^A \otimes \sigma_z^B + \sigma_y^A \otimes \sigma_x^B + \sigma_z^A \otimes \sigma_x^B + \sigma_z^A \otimes \sigma_y^B).$$

Since our choice of spin basis is defined up to an orthogonal transformation, we can use instead of the original spinor basis $|\uparrow\uparrow\rangle, |\uparrow\downarrow\rangle, |\downarrow\uparrow\rangle, |\downarrow\downarrow\rangle$ a singlet-triplet basis obtained with the transformation matrix:

$$O = \begin{pmatrix} 1 & 0 & 0 & 0 \\ 0 & \frac{1}{\sqrt{2}} & 0 & \frac{1}{\sqrt{2}} \\ 0 & \frac{1}{\sqrt{2}} & 0 & -\frac{1}{\sqrt{2}} \\ 0 & 0 & -1 & 0 \end{pmatrix},$$

which leads to the new basis:

$$|S\rangle = \frac{1}{\sqrt{2}}(|\uparrow\downarrow\rangle - |\downarrow\uparrow\rangle), \qquad (18)$$

$$|T_1\rangle = |\uparrow\uparrow\rangle, \quad |T_2\rangle = \frac{1}{\sqrt{2}}(|\uparrow\downarrow\rangle + |\downarrow\uparrow\rangle), \quad |T_3\rangle = |\downarrow\downarrow\rangle.$$

Notice that we have a helicity-dependence in the *Q* and *J*-tensors arising from the off-diagonal components of the dipole moment matrix elements in Eq. (12). With the rotated basis, we obtain the following matrix representation of the interaction energy:

$$W^{(2),aniso} = 2(1_A \otimes 1_B)Tr(Q) + \begin{pmatrix} J_{zz} & (1-i)\sqrt{2}J_0 & -2iJ_0 & 0 \\ (1+i)\sqrt{2}J_0 & 2J_\perp - J_{zz} & -(1-i)\sqrt{2}J_0 & 0 \\ 2iJ_0 & -(1+i)\sqrt{2}J_0 & J_{zz} & 0 \\ 0 & 0 & 0 & -2J_\perp - J_{zz} \end{pmatrix}. \quad (19)$$

This result shows that the singlet and triplet sectors are also decoupled (this even holds for the full expression in Eq. (10), although we only discuss here a simplified case). The diagonalization can be performed analytically and one obtains the following eigenvalues:

$$\lambda_0 = 2Q + 2J_0 + J_{zz},$$
$$\lambda_1 = 2Q - J_{zz} - 2J_0,$$
$$\lambda_{2,3} = 2Q - J_0 + J_\perp \pm \sqrt{8J_0^2 + (J_z - J_0 - J_\perp)^2}.$$

Which eigenvalue has the lowest energy is, however, dependent on the signs of the effective exchange couplings and thus, on the used model parameters. We remark that the corresponding eigenvectors, which are linear combinations of the used spin basis, are also chirality dependent and hence, have different weights depending on whether a hetero- or homo-chiral situation is considered.

An important point to notice is that the *Q*- and *J*-tensors in Eq. (10) identically vanish for a pair of *linear* molecules. In fact, the case of a linear molecule is given by the limiting case of the radius $R_0 \to 0$. Then, it holds:

$$\mu_\alpha^{1m} \mu_\beta^{m1} = \delta_{\alpha z}\delta_{\beta z} 16[1+(-1)^m]^2 (Km)^2 \times \frac{b^2/\pi^4}{(m^2-1)^2} \equiv \delta_{\alpha z}\delta_{\beta z} f(m), \quad (20)$$

from where it follows e.g. for the *K*-tensor that

$$K_{\alpha\beta}^{BA} = -\lambda_{eff}^2 \int_0^\infty \frac{du}{2\pi} \pi^B(iu)\pi^A(iu)\left[\varepsilon_{\alpha\gamma z}\varepsilon_{\beta\gamma' z}T_{\gamma z}T_{\gamma' z}\right] = -\lambda_{eff}^2 \int_0^\infty \frac{du}{2\pi} \pi^B(iu)\pi^A(iu)\left[\varepsilon_{\alpha\gamma z}\varepsilon_{\beta\gamma' z}\delta_{\gamma z}\delta_{\gamma' z}\right] \quad (21)$$

$$= -\lambda_{eff}^2 \int_0^\infty \frac{du}{2\pi} \pi^B(iu)\pi^A(iu)\left[\varepsilon_{\alpha zz}\varepsilon_{\beta zz}\right] \equiv 0.$$

Here, we have defined the functions $\pi^{s=A,B}(iu) = 2\sum_{j\neq 0} f(j)\omega_{j0}^2(s)/u^2 + \omega_{j0}^2(s)$, which depend on *f(j)* defined in Eq. (20). This means, that although the transition dipole matrix does not vanish in the limit of zero radius, the *Q*- and *J*- response tensors will do.

From Eq. (21) it thus follows that the *J*-tensor vanishes. In the second equality in Eq. (21), we have assumed $R = R(1,0,0)$ (the same result will be obtained for any other distance vector as far as it lies on the x-y plane). With a similar analysis one can show that the *Q*-tensor also vanishes. This result suggests that our proposed model in Eq. (1) is not a trivial effect, but it is closely related to the helical symmetry.

We remark, however, that in the case that of a linear molecule and a helical one, the model would in principle yield a non-zero effect, since the corresponding tensors will not necessarily vanish. This case, however, is not of relevance for the discussion, since it does not relate to chirality-dependent interactions. This limitation is related to the fact that the chirality enters into the theory at the level of the zero order Hamiltonian and not in the intermolecular interaction terms of Eq. (1); it could, therefore, be lifted in a more detailed consideration of the symmetries of the interaction terms.

In Figure 1 we show the Lennard-Jones potential from Eq. (13) for different situations. Fig. 1a) displays the energy difference $V_{LJ}(+1,+1,R) - V_{LJ}(+1,-1,R)$ between the homochiral ($s_h^A = s_h^B = 1$) and heterochiral ($s_h^A = 1, s_h^B = -1$) configurations for three representative relative orientations, corresponding to the connecting line between the molecules being along the x-axis $\boldsymbol{R} = R(1,0,0)$, y-axis $\boldsymbol{R} = R(0,1,0)$, and $\boldsymbol{R} = (R/\sqrt{2})(1,1,0)$. The potential has been normalized by the minimum $E_0$ of the London potential and the intermolecular separation by the radius of the helix $R_0$. The energy differences sensitively depend on the orientation, but in all cases and for the used model parameters the homo-chiral case has a lower energy.

A strongly coarse-grained picture of this result would be to think of the spin fluctuations related to the CISS effect as inducing a transient spin polarization, which can be represented by parallel (L-L) or anti-parallel (D-L) pseudo-spin magnetic moments in a pair of molecules, where parallel and anti-parallel have as reference the axis of the helices. In this (oversimplified) picture ferro-like (L-L) and antiferro-like (D-L) interactions could be obtained. Although the sign of the exchange coupling *J* is not obvious *a priori,* it can be assumed to be negative if the homo-chiral situation has a lower energy. In this way, an additional Ising-like pseudo-spin (and thus chirality) dependent contribution to the dispersion energy would be obtained.

Panels b) and c) in Fig. 1 show sections of the individual potential energy curves. While in x-direction both curves have their minima at the same intermolecular distance, there is a small shift in their position for the y-direction (highlighted by the vertical arrows). This indicates a slightly stronger attraction between molecules with the same chirality. This is further highlighted in Fig. 1 d), where we plot the forces corresponding to the homo- (dashed lines) and hetero- (solid lines) chiral cases for three different values of the effective spin-orbit strength (different colors). The zero crossing of the curves indicates the corresponding minimum of the potential. One sees a clear difference between the two cases which becomes more pronounced with increasing $\lambda_{eff}$. Clearly, a more realistic atomistic calculation involving the true helical structure will be ultimately needed to further clarify specific quantitative details of the intermolecular interactions. We also remark that at closer intermolecular separations electronic exchange will become increasingly relevant and may overlap with the effect discussed here, so that in this region it will not be possible to disentangle what is coming from exchange and CISS-related. The theoretical treatment in this shorter range distance may require a different approach such as Symmetry-Adapted Perturbation Theory[72] or the ACFD method. Additionally, terms falling faster than $R^{-6}$ and not discussed here could also contribute at shorter ranges.

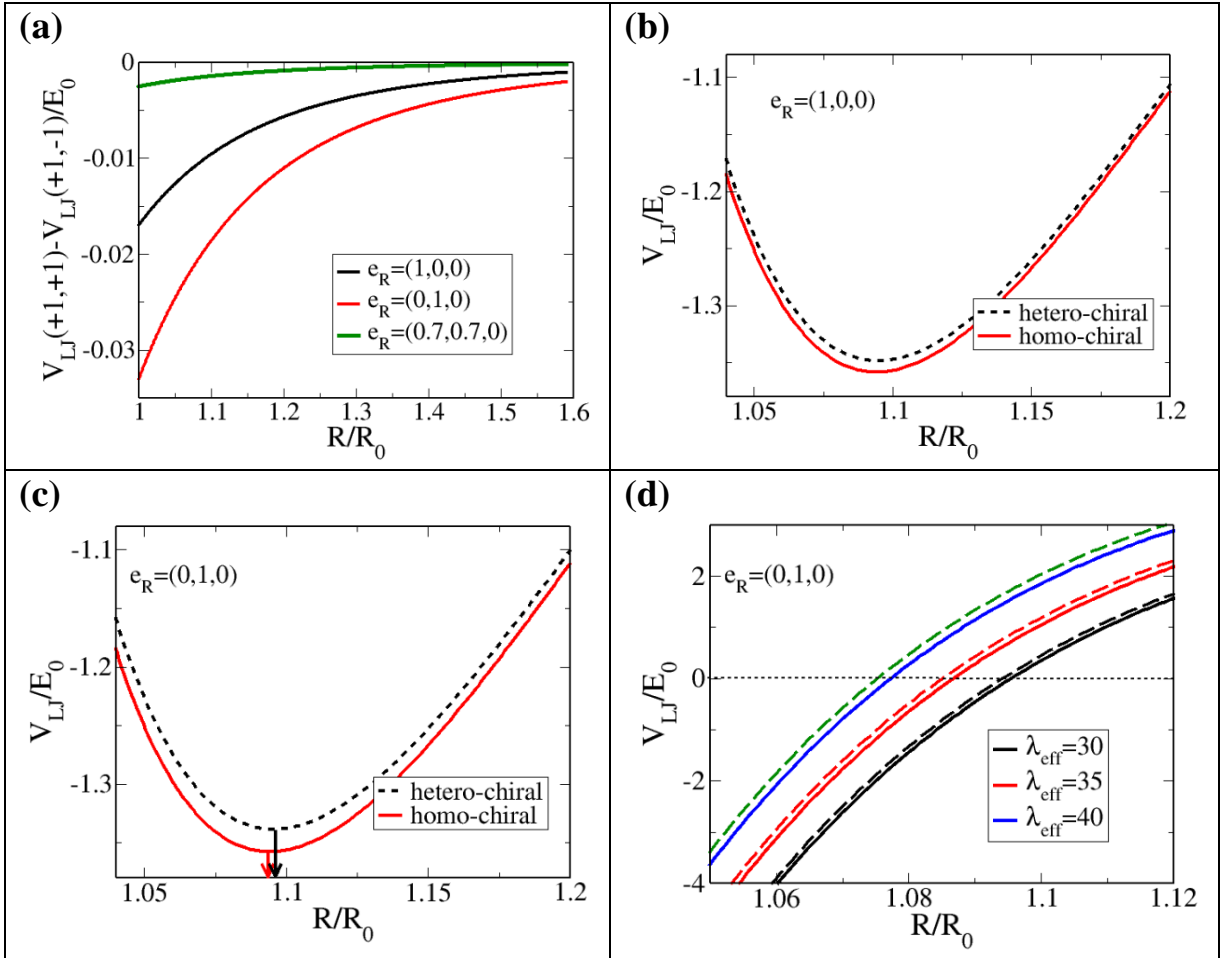

**Figure 1. Influence of the induced intermolecular spin-orbit interaction on the vdW energy in the anisotropic case.** All curves are generated by using the Lennard-Jones potential of Eq. (13). For all cases we have chosen the helical parameters $R_0=b=0.4$ nm and $\lambda_{eff}=30$. Intermolecular separations are scaled with the helix radius and the energy with the minimum of the London energy at the minimum $E_0 = V_{LJ}(R=R_{min}, \lambda_{eff}=0)$. a) Energy difference between homo- and hetero-chiral configurations for three typical orientations with distance vector between the two molecules directed along the x-, y-, and $(1/\sqrt{2})(1,1,0)$-directions. Panels b) and c) show the corresponding potential energy curves for x- and y-directions. The arrows in panel c) highlight the slightly different positions of the minima of the two potential energy curves. Panel d) shows the forces corresponding to hetero-chiral (solid lines) and homo-chiral (dashed lines) interactions, built from the Lennard-Jones potentials for different $\lambda_{eff}$. The zero crossing of each line is the corresponding minimum.

## IV. Beyond the dipole approximation

The model proposed in Eq. (1) is based on the dipole approximation for the field induced by charge fluctuations, and it thus represents the lowest-order term in a multipole expansion of the electrostatic Coulomb interaction. This expansion is, however, not convergent for shorter distances compared with the molecular lengths.[73] We show in this last section a natural extension of the results presented in Eq. (9) to the case where the full electrostatic potential is included.

For this, we first define a charge density fluctuation operator as:

$$\delta\rho_{s=A,B}(\vec{x}) = e\sum_{l=1}^{N_s}\delta(\vec{x}-\vec{x}_l) - \rho_{s,0}(\vec{x}), \tag{22}$$

where $N_{s=A,B}$ is the number of electrons in each molecule and $\vec{x}_l$ is the corresponding position vector.

The electric field induced by this charge density fluctuation acting on a given position $\vec{x}_n$ in the other molecule is then given by:

$$E_\gamma^{s=A,B}(\vec{x}_n) = -\frac{1}{4\pi\varepsilon_0}\frac{\partial}{\partial x_{\gamma n}}\int_{s'\neq s} d\vec{x'}\,\frac{\delta\rho_{s'}(\vec{x'})}{\|\vec{x}_n-\vec{x'}\|}. \tag{23}$$

Finally, the matrix elements of the momentum operator can be written as:

$$\langle n|p_{\alpha,l}(s)|m\rangle = -i\frac{m_e}{e\hbar}\omega_{mn}\langle n|ex_{\alpha,l}(s)|m\rangle = -i\frac{m_e}{e\hbar}\omega_{mn}\int_{s=A,B} d\vec{x}\,\langle n|ex_\alpha\delta\rho_s(\vec{x})|m\rangle \tag{24}$$

$$= -i\frac{m_e}{e\hbar}\omega_{mn}\int_{s=A,B} d\vec{x}\,\langle n|\mu_\alpha^s(\vec{x})|m\rangle.$$

Now, Eq. (3) can be used (we only focus on the terms involving the intermolecular SO, other contributions can be calculated along similar lines) and, after a lengthy but straightforward calculation, the following expression for the second-order energy correction is obtained:

$$\mathbf{W}^{(2)} = \int_B d\vec{x}_1\int_B d\vec{x}_2\,1_A\otimes s_\alpha^B(\vec{x}_1)\Lambda_{BA}^{\alpha\beta}(\vec{x}_1,\vec{x}_2,iu)s_\beta^B(\vec{x}_2) \tag{25}$$

$$+\int_A d\vec{x}_1\int_A d\vec{x}_2\,s_\alpha^A(\vec{x}_1)\Lambda_{AB}^{\alpha\beta}(\vec{x}_1,\vec{x}_2,iu)s_\beta^A(\vec{x}_2)\otimes 1_B$$

$$+(1+P_{AB})\int_B d\vec{x}_1\int_A d\vec{x}_2\,J_{\alpha\beta}^{BA}(\vec{x}_1,\vec{x}_2,iu)s_\alpha^B(\vec{x}_1)\otimes s_\beta^A(\vec{x}_2).$$

Notice that this is still a tensor in spin space. The coupling tensors in the previous expression are defined as follows (in atomic units):

$$\Lambda_{BA}^{\alpha\beta}(\vec{x}_1, \vec{x}_2, iu) = -\lambda_{eff}^2 \varepsilon_{\alpha\kappa\rho} \varepsilon_{\beta\kappa'\rho'} \int_A d\vec{x} \int_A d\vec{y} \int_0^\infty \frac{du}{2\pi}$$

$$\times \left( \frac{\partial}{\partial x_{1\rho}} \frac{1}{\|\vec{x}_1 - \vec{x}\|} \right) \left( \frac{\partial}{\partial x_{2\rho'}} \frac{1}{\|\vec{x}_2 - \vec{y}\|} \right) \Gamma_B^{\kappa\kappa'}(\vec{x}_1, \vec{x}_2, iu) \alpha_A(\vec{x}, \vec{y}, iu), \quad (26)$$

$$\Gamma_s^{\kappa\kappa'}(\vec{y}_1, \vec{y}_2, iu) = 2 \sum_{n \neq 0} \frac{\omega_{n0}^3(s)}{\omega_{n0}^2(s) + u^2} \langle 0 | x_{1,\kappa} \delta\rho_s(\vec{y}_1) | n \rangle \langle n | x_{2,\kappa'} \delta\rho_s(\vec{y}_2) | 0 \rangle,$$

$$\alpha_s(\vec{y}_1, \vec{y}_2, iu) = 2 \sum_{n \neq 0} \frac{\omega_{n0}(s)}{\omega_{n0}^2(s) + u^2} \langle 0 | \delta\rho_s(\vec{y}_1) | n \rangle \langle n | \delta\rho_s(\vec{y}_2) | 0 \rangle.$$

The exchange tensor $J_{BA}^{\alpha\beta}(\vec{x}_1, \vec{x}_2, iu)$ is given by:

$$J_{BA}^{\alpha\beta}(\vec{x}_1, \vec{x}_2, iu) = -\lambda_{eff}^2 \varepsilon_{\alpha\kappa\rho} \varepsilon_{\beta\kappa'\rho'} \int_A d\vec{x} \int_B d\vec{y} \int_0^\infty \frac{du}{2\pi} \quad (27)$$

$$\times \left( \frac{\partial}{\partial x_{1\rho}} \frac{1}{\|\vec{x}_1 - \vec{x}\|} \right) \Pi_A^\kappa(\vec{x}, \vec{x}_2, iu) \left( \frac{\partial}{\partial x_{2\rho'}} \frac{1}{\|\vec{x}_2 - \vec{y}\|} \right) \Pi_B^\kappa(\vec{y}, \vec{x}_1, iu)],$$

$$\Pi_A^\kappa(\vec{x}, \vec{x}_2, iu) = 2 \sum_{n \neq 0} \frac{\omega_{n0}^2(s)}{\omega_{n0}^2(s) + u^2} \langle 0 | \delta\rho_A(\vec{x}) | n \rangle \langle n | x_{2,\kappa} \delta\rho_A(\vec{x}_2) | 0 \rangle,$$

$$\Pi_B^\kappa(\vec{x}_1, \vec{y}, iu) = 2 \sum_{m \neq 0} \frac{\omega_{m0}^2(s)}{\omega_{m0}^2(s) + u^2} \langle 0 | x_{1\kappa} \delta\rho_B(\vec{x}_1) | m \rangle \langle m | \delta\rho_B(\vec{y}) | 0 \rangle$$

The tensor $\Lambda_{BA}^{\alpha\beta}(\vec{x}_1, \vec{x}_2, iu)$ can be interpreted as a non-local pseudo-magnetic field acting on each of the molecules and correlates dipole fluctuations and charge density fluctuations via the $\Gamma$ and $\alpha$ response functions. The non-local exchange tensor $J_{BA}^{\alpha\beta}(\vec{x}_1, \vec{x}_2, iu)$ connects charge fluctuations on different molecules: under the influence of the intermolecular electrostatic field $\vec{E}_{AB}$, the $\Pi$-tensors describe the propagation of a density fluctuation from a point $\vec{y}_1$ on one molecule to another point $\vec{y}_2$ on the same molecule. This fluctuation propagates back via the field $\vec{E}_{AB}$ to the other molecule, closing the loop (due to charge conservation). Eq. (25) generalizes the results of the previous sections to intermediate intermolecular separation regions (but still in the region where wave function overlap can be safely neglected) and also allows, in principle, for a description of extended molecular structures. The results of Sec. III can be recovered by performing a multipole expansion of the electrostatic fields and keeping only the dipole contributions.

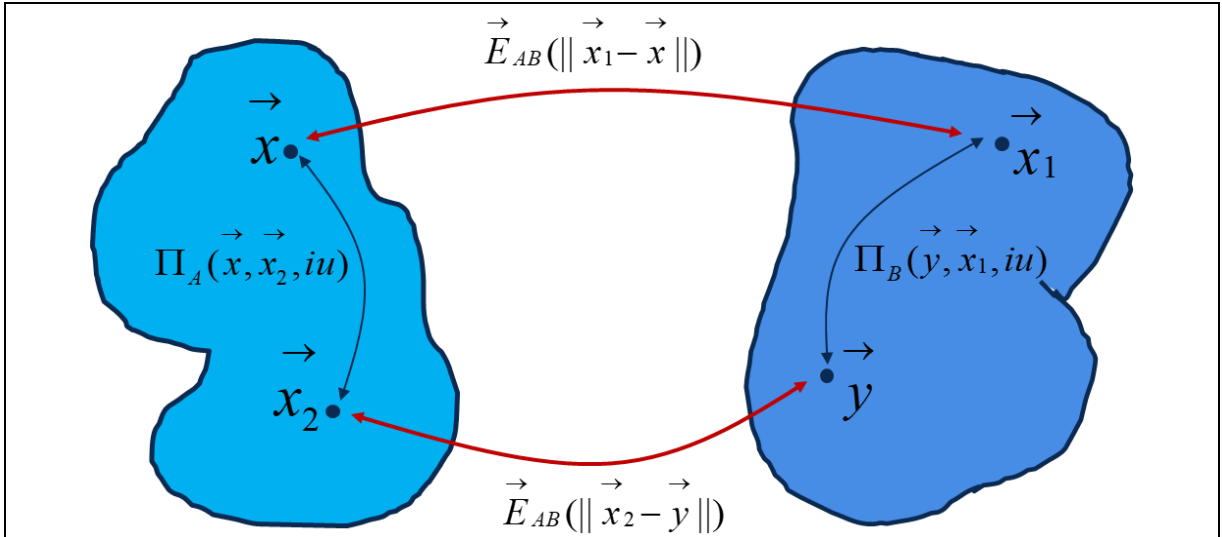

**Figure 2.** Schematic representation of the structure of the exchange tensor $J_{BA}^{\alpha\beta}(\vec{x}_1, \vec{x}_2, iu)$ from Eq. (27). The nonlocal response functions $\Pi_B$ and $\Pi_A$ connect charge fluctuations at different points within the same molecule. The intermolecular coupling is mediated by the electrostatic fields $\vec{E}_{AB}$.

## V.  CONCLUSIONS

In conclusion, we have discussed the consequences of assuming the presence of mutually induced spin-orbit coupling in helical molecules in the near-zone range where electronic exchange effects can be neglected. Our results hint at a spin-dependent correction to the standard London dispersion forces, which has the form of a generalized pseudo spin-spin exchange interaction term scaling with $R^{-6}$, in a similar way as the London dispersion energy. This generic term encodes the interplay between charge fluctuations (hidden in the effective exchange coupling tensor **J**) and spin-dependent effects, and provides a phenomenological way of accounting for CISS-related processes. The fact that transient spin fluctuations are responsible for the effect implies that no time-reversal symmetry breaking is needed as in the case of spin-polarized electron transport and electron transfer.[74]

In the light of the results obtained here, it can be speculated whether the spin-dependent correction terms to the dispersion interaction may also provide an additional contribution –besides the intrinsic spin-orbit terms commonly assumed in the models so far– to the description of the CISS effect in self-assembled monolayers, where intermolecular interactions cannot be neglected and the relative molecular orientations are nearly fixed in space.

We remark that in general terms, besides the proposed spin-orbit mediated dispersion terms and the electric dipole-dipole interactions leading to the London dispersion, there are also magnetic dipole-dipole interactions, which should be included in a complete calculation of dispersion forces for chiral molecules.[9,58]

Concerning possible corrections to chirality discriminating intermolecular interactions involving the rotatory power (RP),[10,14,58] one may wonder if, within 2nd order perturbation theory, mixed SOC-magnetic terms may yield additional corrections to these terms, i.e. contributions with the generic form $\langle H_{SO} \rangle \langle H_{dip-dip}^{mag} \rangle$. Such terms would also involve products of transition matrix elements of electrical and magnetic dipole moments, similar to the RP. A very preliminary calculation shows that, indeed, a possible correction to the RP-related terms[10,14,58] can play a role and it scales roughly as (in atomic units)

$\alpha^2 \lambda_{eff} \langle \omega \rangle$ with $\alpha = 1/137$ being the fine structure constant, and $\langle \omega \rangle$ is a typical excitation energy scale. Since the standard RP contribution scales as $\alpha^2$,[9] the leading correction to it will be proportional to $\lambda_{eff} \langle \omega \rangle$. Thus, a large excitation energy from the ground state may provide a non-negligible correction to the RP-terms. An in-depth study of this issue will be reported elsewhere.

Finally, our results suggest that classical force field descriptions of van-der Waals interactions, commonly used in atomistic Molecular Dynamics simulations, may require additional modifications to include the effects discussed here.[21] We believe that our approach yields a relatively simple way to extend typical force fields to include spin-dependent effects.

## SUPPLEMENTARY MATERIAL

The Supplementary Material includes few additional details of the analytical calculations presented in the main text as well as details of the Density-Functional-Theory calculations on helicene enantiomers.

## ACKNOWLEDGMENTS

The authors thank E. Diaz and F. Dominguez-Adame for very useful discussions. A.D. and G. C. acknowledge financial support from the Volkswagen Stiftung within the project *Spintronic Components based on Chiral Molecules – From single to multiple elements –* (grant No. 88366). This work was also supported by the German Research Foundation (DFG) within the project *Theoretical Studies on Chirality-Induced Spin Selectivity* (CU 44/55-1) and by the transCampus Research Award *Disentangling the Design Principles of Chiral-Induced Spin Selectivity (CISS) at the Molecule/Electrode Interface for Practical Spintronic Applications* (tC RA 2020-01). V. M. acknowledges a Fellowship from Ikerbasque, the Basque Foundation for Science (Spain). After finishing this study, we became aware of related work by Per Hedegård, which addresses the problem from a different perspective by considering intramolecular spin-orbit coupling in interplay with intermolecular dipole-dipole interactions. We thank Per Hedegård for drawing our attention to his work.

## DATA AVAILABILITY

The data that support the findings of this study are available from the corresponding author upon reasonable request.

# Supplementary Material

# The contribution of intermolecular spin interactions to the London dispersion forces between chiral molecules

M. Geyer,[†] R. Gutierrez,[†] V. Mujica,[‡] J. F. Rivas Silva,[††] A. Dianat,[†] G. Cuniberti[†,¶]

[†] Institute for Materials Science and Max Bergmann Center of Biomaterials, Dresden University of Technology, 01062 Dresden, Germany

[‡] Arizona State University, School of Molecular Sciences, PO Box 871604, Tempe, Arizona 85287-1604, USA

[¶] Dresden Center for Computational Materials Science and Center for Advancing Electronics Dresden, TU Dresden, 01062 Dresden, Germany

[††] Instituto de Física Luis Rivera Terrazas, Benemérita Universidad Autónoma de Puebla, Apdo. Postal J48, Col. San Manuel, Puebla, Pue. C. P. 72570, Mexico

## 1. Isotropic limit

*London dispersion energy.* As mentioned in Sec. II.D. in the main text, we use an approximate expression $C_{London}^{iso}$ to describe the standard London part of the dispersion potential. Although this is not compelling, it helps to isolate the helicity dependence in the indSO-dependent terms and to better clarify their contributions.

The standard London result for the dispersion interactions in second-order perturbation theory is given by:[1]

$$W_{London}^{(2)} = -\frac{1}{(4\pi\varepsilon_0)^2} T_{\gamma\rho} T_{\gamma'\rho'} \sum_{n,m\neq 0} \frac{1}{\omega_{n0} + \omega_{m0}} \mu_\gamma^{0n}(A)\mu_{\gamma'}^{n0}(A)\mu_\rho^{0m}(B)\mu_{\rho'}^{m0}(B) \quad \text{(S1)}$$

$$= -\frac{1}{2\pi}\frac{1}{(4\pi\varepsilon_0)^2} T_{\gamma\rho} T_{\gamma'\rho'} \int_0^\infty du\, \alpha_{\gamma\gamma'}^A(iu)\alpha_{\rho\rho'}^B(iu) = -\frac{1}{2\pi}\frac{1}{(4\pi\varepsilon_0)^2}\int_0^\infty du\, \text{Tr}\{\boldsymbol{T}\boldsymbol{\alpha}^A(iu)\boldsymbol{T}\boldsymbol{\alpha}^B(iu)\},$$

If the polarizability tensors are isotropic, then it holds:

$$\alpha_{\rho\rho'}^{j=A,B}(iu) = \delta_{\rho\rho'}\frac{2}{3}\sum_{n\neq 0}\frac{\omega_{n0}}{u^2+\omega_{m0}^2}\mu_\rho^{0n}(j)\mu_\rho^{n0}(j) = \delta_{\rho\rho'}\frac{2}{3}\sum_{n\neq 0}\frac{\omega_{n0}}{u^2+(\omega_{n0})^2}|\langle 0|\boldsymbol{\mu}^{j=A,B}|n\rangle|^2 . \quad \text{(S2)}$$

For a fixed relative spatial orientation of the two molecules and without loss of generality (due to the isotropy of the problem, there is no preferred orientation), we can choose $\vec{R} = (R,0,0)$ to obtain:

$$T = \frac{1}{R^3} \begin{pmatrix} -2 & 0 & 0 \\ 0 & 1 & 0 \\ 0 & 0 & 1 \end{pmatrix}.$$

Using the isotropic approximation and introducing atomic units, we get:

$$W^{(2)}_{London}(R) = -\frac{12}{\pi} \frac{1}{R^6} \times \int_0^\infty du \; \alpha^A(iu)\alpha^B(iu) = -\frac{C^{iso}_{London}}{R^6} = -\frac{C_6}{R^6}. \tag{S3}$$

Notice that the influence of the helicity would appear only in the off-diagonal components of the susceptibility tensors by using the helix model of Sec. II.C. Hence, within the isotropic approximation only a dependence on radius and pitch will survive, but no helicity dependence. To compute the corresponding polarizability functions $\alpha^A(iu)$, we will adopt the following expression for the product of the dipole matrix elements:

$$\mu_\alpha^{1m} \mu_\beta^{m1} = 16[1+(-1)^m]^2 (Km)^2 \times \tag{S4}$$

$$\frac{1}{3}\left\{ \frac{4R_0^2/\pi^2}{[(m-1)^2 - 4K^2]^2[(m+1)^2 - 4K^2]^2} + \frac{b^2/\pi^4}{(m^2-1)^2} \right\} \times Diag(0,1,1),$$

$$\mu_\alpha^{1m} \mu_\beta^{m1} = \delta_{\alpha\beta} \frac{R^2}{12} \text{ for } m = 2K \pm 1.$$

*Degenerate perturbation theory in the isotropic case*. Since we are using for the sake of simplicity the isotropic limit to estimate the ratio between London and indSO corrections, we delineate here the isotropic approximation for Eq. (10). Assuming that all susceptibility tensors have the same structure as in Eq. (S2), the *Q*- and *K*-tensors simplify to:

$$Q^{BA}_{\alpha\beta} = -\lambda_{eff}^2 \frac{1}{R^6} \int_0^\infty \frac{du}{2\pi} \; \Gamma^B(iu)\alpha^A(iu)\left[\delta_{\alpha\beta} R^6 Tr(T^2) - R^6 T^2_{\alpha\beta}\right], \tag{S5}$$

and

$$K^{BA}_{\alpha\beta} = -2\lambda_{eff}^2 \frac{1}{R^6} \int_0^\infty \frac{du}{2\pi} \; \Pi^B(iu)\Pi^A(iu)\left[\delta_{\alpha\beta} + R^3 T_{\alpha\beta}\right]. \tag{S6}$$

Assuming the same orientation as before, $\vec{R} = (R,0,0)$, we obtain:

$$Q(R) = -\lambda_{eff}^2 \frac{1}{R^6} \int_0^\infty \frac{du}{2\pi} \; \Gamma(iu)\alpha(iu), \tag{S7}$$

$$K(R) = -\lambda_{eff}^2 \frac{1}{R^6} \int_0^\infty \frac{du}{2\pi} \; \Pi(iu)\Pi(iu). \tag{S8}$$

With these simplification, Eq. (10) becomes for the same type of molecules:

$$W^{(2)}(R) = 24Q(R)(1_A \otimes 1_B) + 4K(R)\left(-\sigma_x^A \otimes \sigma_x^B + 2\sigma_y^A \otimes \sigma_y^B + 2\sigma_z^A \otimes \sigma_z^B\right). \tag{S9}$$

Using the same singlet-triplet basis as in the main text, we can diagonalize Eq. (S9) in spin space and obtain the ground state energy. This is the expression we use then in Eq. (14) in the main text.

## 2. Density-functional theory (DFT) based calculations on helicene enantiomers: energy scales

The model proposed in the main text is largely phenomenological. We have attempted to provide a very rough estimate of the possible orders of magnitude of the coupling constant $\lambda_{SO}$ in Eq. (1) based on a series of DFT calculations that we are presenting here. Still, a realistic estimate remains open for the moment, since there is no available DFT-based code where the proposed spin-dependent corrections to the dispersion energy are included.

The goal of the DFT calculations presented below is rather to provide an insight into the typical energy scales involved in the dispersion interaction as well as the influence of *intra-molecular* spin-orbit interactions on the dispersion energy.

We have chosen Helicene as a structurally simple carbon-based model molecular system, which does not have a sizeable permanent dipole moment in the ground state, so that the main contribution to vdW interactions is expected to arise from dispersion interactions. We have then considered the energetics of dispersion for D-D as well as L-D pairs.

*Case 1.* In a first approach, calculations with the GAUSSIAN code[6] were carried out, allowing for a full optimization of the intermolecular distance and relative orientation of a molecule pair (either D-D or L-D helicenes), as shown in Fig. **S1**. Based on the specular symmetry displayed by the molecular pair, we denote by SPEC the case which presents equal specular positions of the atoms of molecules of the pair, considering a plane perpendicular to the helix axis. This corresponds to *different* (L-D) enantiomers. The second arrangement, denoted by ANTISPEC, presents positions on a left-right specular symmetry and corresponds to the *same* enantiomers (L-L or D-D). The DFT calculations were carried out using a Dunning atomic basis (cc-pvdz) and the W97XD hybrid functional from Chai et al..[7] This functional is based on a B97 functional with a convenient treatment of long- and short-range effects of the exchange interaction, and a special term taking into account dispersive forces by using a Grimme-type term. It is important to highlight that in all cases a full structural optimization of the corresponding molecular pair was carried out. In Fig. **S1** the respective equilibrium geometries are presented, the color code on the atoms relating to their Mulliken charge values (red negative, green positive). In Table **S1,** we summarize the obtained total energies and dipole moments of the SPEC and ANTISPEC configurations.

**Table S1.** Total interaction energies and dipole moments of the homo (ANTISPEC)- and hetero (SPEC) chiral helicene conformations.

| Case | Total energy (in a.u.) | Dipole moment (in D) |
|---|---|---|
| SPEC | -2000.37298483 | 0.626 |
| ANTISPEC | -2000.35710265 | 0.935 |

From Table **S1** it becomes evident that the energy difference $\Delta E_{ch} = E_{DL} - E_{DD}$, which can be considered as a measure of chiral discrimination, has a sizeable value 0.353 eV (8.134 kcal/mol). *To gain further insight into the role of the dispersion interaction, we performed similar calculations without including dispersion forces in DFT and using the B3LYP functional.[8] We obtained evidence that there is almost no sizeable difference in the total energy of homo- and hetero-chiral pairs, yielding now* $\Delta E_{ch} = 27$

*meV, which is roughly a factor 10 smaller than the previously obtained value of 0.353 eV and of the order of room temperature.*

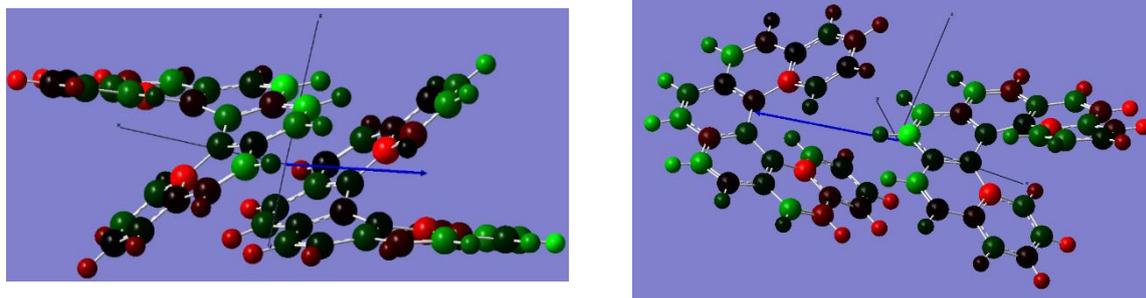

**Figure S1**. Optimized geometry of a pair of helicenes in the *heterochiral* (SPEC, left panel) and *homochiral* (ANTISPEC, right panel) configurations. Different colors label Mulliken charges.

To further illustrate our results, we use the NCI (Non Covalent Interaction Index) method by using the *Multiwfn* code,[9,10] which makes an analysis of the surface of the electronic density $\rho$ of a molecular system. The eigenvalues ($\lambda_1$, $\lambda_2$, $\lambda_3$) of the Laplacian $\Delta\rho$ characterize the behavior according to the sign of the eigenvalues. Eigenvalues with one positive and two negative signs ($\lambda_1<0$, $\lambda_2<0$, $\lambda_3>0$) characterize particularly the bonding at the interatomic regions between atoms. The value of the reduced density gradient (RDG) $\sigma$, coming from the density and its first derivative, $\sigma = 1/(2(3\pi^2)^{1/3}|\nabla\rho|/\rho^{4/3}$ is obtained versus the product of sign($\lambda_2$)$\rho$. Color code is added to better illustrate the type of bonding found in different regions: negative values of sign($\lambda_2$)$\rho$ indicate weak attraction on the red side, around zero value, the presence of Van der Waals interactions is shown with green, and steric forces on blue. We present here the Fig. **S3** for the function $\sigma$ depicted over $\rho$, and Fig. **S4** for the numerical value of $\sigma$ vs. sign($\lambda$2)$\rho$, for ANTISPEC and SPEC cases, respectively.

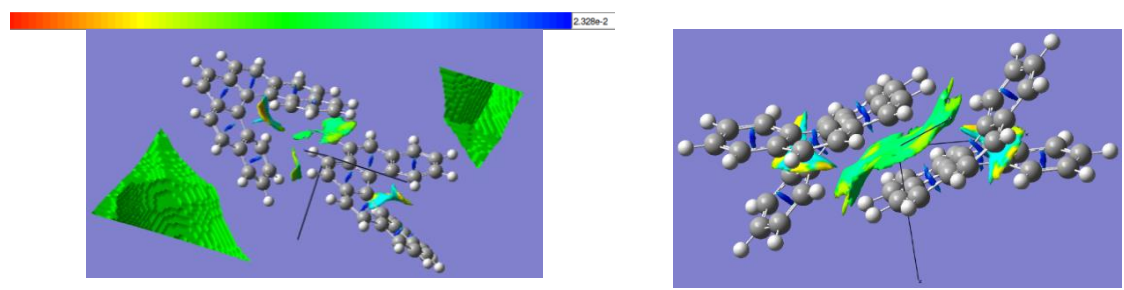

**Figure S3.** NCI analysis of ANTISPEC case (left) and SPEC (right) cases, showing the $\sigma$ function over density in a contour map.

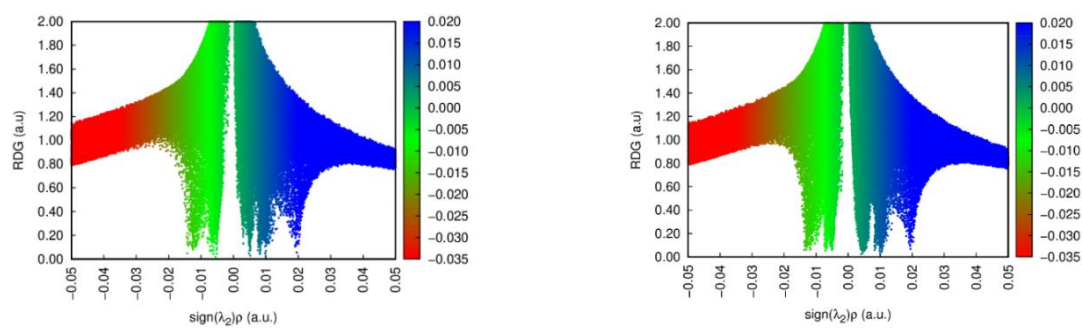

**Figure S4**. NCI analysis of ANTISPEC case (left panel) and SPEC case (right panel), showing the $\sigma$ function versus sign($\lambda_2$)$\rho$.

The difference between the homo-chiral and hetero-chiral situations can be clearly seen in these graphs Fig. **S3**, especially at the middle region between the two helicene molecules the SPEC case displays a major contribution of vdW interactions, explaining the lower energy of the optimized configuration when compared with the ANTISPEC case. It can also be seen that in both cases other types of forces also appear, highlighted e.g. as the blue zones at the middle of the bencene rings. Fig. **S4** show the ranges of variables discussed above.

These calculations illustrate the well-known fact that DFT with added dispersion corrections is able to discriminate energetically between equal and different chiralities. Although at the range of equilibrium distances (of the order of 3.0 Å) obtained upon the geometry optimization electronic exchange interactions still play a role, we have found out that the major contribution to the discriminative interactions is provided by the dispersion interactions. The obtained energy differences L-L vs. D-L of the order of 350 meV are a result of purely configurational effects, i.e. the ground state orientations of L-L and D-L pairs are quite different.

*Case 2.* While the previous calculations include a full optimization of inter-molecular distances and relative orientations, the case we have studied in our model system in the main text is more restrictive and involves two helices with their helical axis parallel to each other. It is, therefore, of interest to check how the energetics of dispersion interactions is modified in this more restricted configuration. We then consider the situation of Figure **S5**, where now only structural optimization of the molecular structures at a given separation of roughly 3.5 Å was carried out. While these configurations may not be local minima on the global potential energy surface, they are closer to the model situation and serve to provide orders of magnitude of the energy scales.

We computed again the energy differences between D-D and D-L conformations of helicene using the Vienna *ab initio* simulation package (VASP) with the augmented-plane-wave (PAW) method and the GGA-PAW-PBE XC-functional. An energy cutoff of 400 eV was used. Dispersion forces were now accounted for at three different levels: Grimme D2 parametrization,[2] Grimme D3,[3] and the Tkatchenko-Scheffler (TS) approach.[4]

Although D2 and D3 parametrizations account for modifications of the local atomic environment in comparison with the case of free atoms, they do not depend on the charge density redistribution. This is the advantage of the TS approach. The obtained energy differences between enantiomer configurations $\Delta E_{ch} = E_{DD} - E_{DL}$ are shown in Table **S2**. Notice that $\Delta E_{ch}$ at the TS level is roughly of the same order as the value obtained in *Case 1* above with full optimization. Switching off the dispersion interactions considerably reduces $\Delta E_{ch}$. Notice also that using the Grimme D3 level, which includes $R^{-8}$ contributions leads to a quantitative improvement when compared with D2. This might be an indication that quadrupole contributions are playing an important role in this situation. This is also a known fact from molecular Quantum Electrodynamics calculations for chiral molecules.[5]

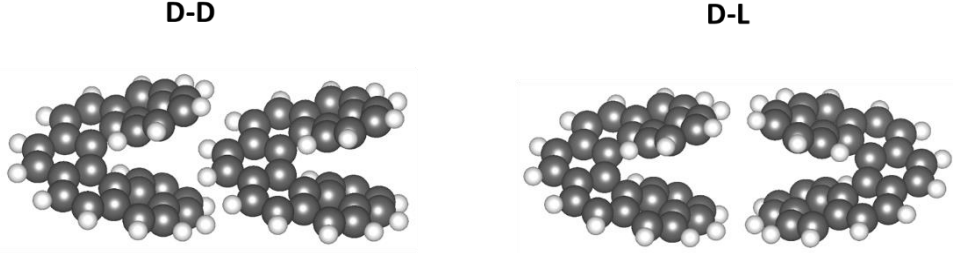

**Figure S5**. Arrangement of hexahelicene enantiomers used for the DFT calculations in *Case 2*. The helical axes of the molecules were kept parallel to each other and no further optimization of the relative distance or orientation was performed.

**Table S2**. Absolute energy difference $\Delta E_{ch} = |E_{DD} - E_{DL}|$ (in meV) between D-D and D-L enantiomer configurations. The corresponding molecular arrangements are shown in Figure **S5**.

| | |
|---|---|
| No dispersion | 45 |
| D2 | 56 |
| D3 | 160 |
| Tkatchenko-Scheffler (TS) | 240 |

*Case 3*. The results presented so far illustrate the typical orders of magnitude associated with different treatments of dispersion. We have, however, not included in any case spin-orbit interactions. In this last case, we address the changes in dispersion energies when considering *intramolecular spin-orbit* coupling. The calculations here were also carried out with the VASP code, see *Case 2* for the technical details. Additionally, we have considered few intermolecular separations. The results are summarized in the following Tables **S3-S7.** We only use the TS approximation to dispersion, since it depends on intramolecular charge reorganization and can, therefore, better catch eventual changes in the energetics when comparing cases with and w/o intramolecular SO interactions.

In the tables, we indicate the total energy of the corresponding molecular pair $E_{tot}$ and the total contribution from dispersion interactions $E_{disp}$ (including both intra- and inter-molecular components). Moreover, we have also computed two other quantities, which provide an estimate for <u>intermolecular</u> contributions:

- A pseudo-binding energy $E_{int/D} = E_{tot} - 2 \times (E_{tot}^{single} - E_{disp}^{single})$, where we subtract from the total energy two times the non-dispersive part of the total energy of an isolated molecule.
- The intermolecular component of the dispersion energy $E_{disp}^{int} = E_{disp}^{tot} - 2 \times E_{disp}^{single}$, where we subtract from the total dispersion two times the (purely intramolecular) dispersion energy of an isolated molecule.

Notice that with the used definition of $E_{int/D}$, $E_{int/D}(R \to \infty) \to 2 \times E_{disp}^{intra} \approx const.$, which is just the intramolecular dispersion energy. Then, the ratio $E_{disp}^{int} / E_{int/D}$, which provides a measure of the relative contribution of intermolecular dispersion to the intermolecular interaction energy will converge to zero in a smooth way. This, of course, is not the only way of quantifying the different contributions and other combinations may be considered as well.

**Tables S3 and S4** show the DFT-based energetics of the **homochiral (L-L) configuration without (Table S3)** and **with (Table S4)** <u>intra-molecular</u> spin-orbit (SO) interactions. Energies are given in eV.

| | | | | | |
|---|---|---|---|---|---|
| 3,73 | -676,1421 | -2,7044 | -2,7305 | -0,3117 | 0,1142 |
| 3,97 | -676,1469 | -2,7126 | -2,7354 | -0,3199 | 0,1169 |
| 4,2  | -676,1232 | -2,6823 | -2,7116 | -0,2896 | 0,1068 |
| 4,6  | -676,0684 | -2,6572 | -2,6569 | -0,2645 | 0,0996 |
| **4,9** | -675,9911 | -2,5648 | -2,5796 | -0,1721 | 0,0667 |
| 5,2  | -675,9030 | -2,4663 | -2,4914 | -0,0736 | 0,0295 |

Table S4. DFT-based total energy ($E_{tot}$), dispersion part ($E_{disp}$) of the total energy, pseudo-interaction energy ($E_{int/D}$), intermolecular dispersion part ($E_{disp}^{int}$), and the ratio of the two latter.

| R/Å | $E_{tot}$ | $E_{disp}$ | $E_{int/D}$ | $E_{disp}^{int}$ | $E_{disp}^{int}/E_{int/D}$ |
|---|---|---|---|---|---|
| 3,73 | -676,1508 | -2,7178 | -2,7375 | -0,3252 | 0,1188 |
| 3,97 | -676,1438 | -2,7322 | -2,7305 | -0,3396 | 0,1244 |
| **4,2** | -676,4583 | -2,7260 | -3,0451 | -0,3334 | 0,1095 |
| 4,6 | -675,9967 | -2,5467 | -2,5835 | -0,1541 | 0,0597 |
| **4,9** | -675,9834 | -2,5536 | -2,5701 | -0,1610 | 0,0626 |
| 5,2 | -675,8954 | -2,4587 | -2,4822 | -0,0661 | 0,0266 |

For the case without intra-molecular SO, Table **S3**, the ratio $E_{disp}^{int}/E_{int/D}$ becomes as expected smaller with increasing separation. Note that the total energy displays a local shallow minimum around 4.2 Å which can clearly be associated with pure dispersion effects.

Similarly, upon including intra-molecular SO, Table **S4,** the ratio $E_{disp}^{int}/E_{int/D}$ decays again smoothly, but the total energy has now, besides the local minimum around 4.2 Å, a another very shallow one in the region around 4.9 Å.

**Tables S5 and S6** show DFT-based energetics of the **heterochiral (L-D)** case **without (Table S5)** and **with (Table S6)** intra-molecular spin-orbit (SO) interactions. Energies are given in eV.

Table **S5**. DFT-based total energy ($E_{tot}$), dispersion part ($E_{disp}$) of the total energy, pseudo-interaction energy ($E_{int/D}$), intermolecular dispersion part ($E_{disp}^{int}$), and the ratio of the two latter.

| R/Å | $E_{tot}$ | $E_{disp}$ | $E_{int/D}$ | $E_{disp}^{int}$ | $E_{disp}^{int}/E_{int/D}$ |
|---|---|---|---|---|---|
| 3,73 | -675,9660 | -2,6347 | -2,5545 | -0,2420 | 0,0947 |
| 4,03 | -675,9841 | -2,6022 | -2,5726 | -0,2095 | 0,0814 |
| **4,35** | -675,9531 | -2,5546 | -2,5416 | -0,1619 | 0,0637 |
| 4,63 | -675,9066 | -2,5088 | -2,4950 | -0,1161 | 0,0465 |
| **4,93** | -675,9964 | -2,6283 | -2,5849 | -0,2356 | 0,0911 |
| 5,2 | -675,8573 | -2,4617 | -2,4458 | -0,0690 | 0,0282 |

Table **S6**. DFT-based total energy ($E_{tot}$), dispersion part ($E_{disp}$) of the total energy, pseudo-interaction energy ($E_{int/D}$), intermolecular dispersion part ($E_{disp}^{int}$), and the ratio of the two latter.

| R/Å | $E_{tot}$ | $E_{disp}$ | $E_{int/D}$ | $E_{disp}^{int}$ | $E_{disp}^{int}/E_{int/D}$ |
|---|---|---|---|---|---|

| | | | | |
|---|---|---|---|---|
| 3,73 | -675,9664 | -2,6347 | -2,5531 | -0,2421 | 0,0948 |
| 4,03 | -675,9881 | -2,6051 | -2,5748 | -0,2125 | 0,0825 |
| **4,35** | -676,0219 | -2,6235 | -2,6087 | -0,2309 | 0,0885 |
| 4,63 | -676,0123 | -2,6208 | -2,5990 | -0,2282 | 0,0878 |
| 4,93 | -675,9920 | -2,6180 | -2,5788 | -0,2254 | 0,0874 |
| 5,2 | -675,8550 | -2,4593 | -2,4417 | -0,0667 | 0,0273 |

Here the behavior is less smooth due to stronger structural relaxation effects of the heterochiral pair. Without SO, Table **S5**, there are two shallow minima around 4,35 Å and 4,93 Å, while upon inclusion of SO, Table **S6**, only the shallow minimum around 4,35 Å was obtained. The stronger relaxation effects are reflected also in the non-monotonous behavior of $E_{disp}^{int}/E_{int/D}$ both with and w/o SO. Clearly, a smoother energy profile can be obtained with a finer discretization of the separation axis, but more detailed information is not relevant for our goal of getting typical orders of magnitude.

What we can now compare is the difference $E_{int/D}^{SO} - E_{int/D}^{no-SO}$ of the dispersion contribution to the interaction energy with and w/o SO for L-L and L-D pairs. This quantity gives a rough order of magnitude insight into the influence of <u>intramolecular SO</u> on <u>intermolecular dispersion</u> interactions. The results are shown in Table **S7.** For completeness we also show $E_{int/D}^{SO} - E_{int/D}^{no-SO}$. In general, there is no smooth trend in the energy differences with an abrupt increase around the regions where shallow local minima were found. Therefore, the strongest effects seem to appear as a result of the interplay between structural relaxation and the related charge density fluctuations, the latter affecting on its turn the SO interaction. Test calculations with the D2 approximation (not shown here) showed clearer that the strongest change in dispersion energy upon switching SO took place around the local vdW minimum. The last column of Table **S7** shows the parameter $\eta = |(E_{disp}^{int,SO}/E_{int/D}^{SO}) - (E_{disp}^{int,no-SO}/E_{int/D}^{no-SO})|$, which is the difference in the ratios $E_{disp}^{int}/E_{int/D}$ with and w/o SO. This ratio, as mentioned above, is an approximate measure of the contribution of dispersion to the intermolecular interaction energy. We see that the strongest change in $\eta$ takes place around 4,6 Å. These separations are well within the vdW regime. In general, a more detailed inquiry may be needed but it goes beyond the scope of these calculations, whose only purpose was to highlight the sensitivity of dispersion to structural and electronic (SOC) effects.

Table **S7**. Differences in the interaction part of the total energy $|E_{int/D}^{SO} - E_{int/D}^{no-SO}|$ with and w/o SO interaction as well as difference in the dispersion part of the total dispersion $|E_{disp}^{int,SO} - E_{disp}^{int,no-SO}|$ for homo (L-L or D-D) and hetero (L-D) helicene molecular pairs using the Tkatshenko-Scheffler approximation for the dispersion interactions. The parameter $\eta$ is defined in the text. Energies are in meV.

| Enantiomer pair | R/Å | $\|E_{int/D}^{SO} - E_{int/D}^{no-SO}\|$ | $\|E_{disp}^{int,SO} - E_{disp}^{int,no-SO}\|$ | $\eta$ |
|---|---|---|---|---|
| **L-L or D-D** | | | | |
| | 3,73 | 7,0 | 13,5 | 0,00463 |
| | 3,97 | 4,9 | 19,7 | 0,00743 |
| | 4,2 | 333,4 | 43,8 | 0,00267 |
| | 4,6 | 73,4 | 110,4 | 0,039893 |
| | 4,9 | 9,5 | 11,2 | 0,00410 |
| | 5,2 | 9,2 | 7,5 | 0,00291 |
| **L-D** |  |  |  |  |
|  | 3,73 | 6.15 | 0,09 | 8,48866E-05 |
|  | 4,03 | 1,3 | 2,94 | 0,00107 |
|  | 4,35 | 2,2 | 68,99 | 0,024808 |
|  | 4,63 | 67,1 | 112,07 | 0,041258 |
|  | 4,93 | 104,0 | 10,18 | 0,00373 |
|  | 5,2 | 6,1 | 2,36 | 0,000919 |

As already mentioned, these DFT results do not directly provide a way to parametrize the interaction strength in Eq. (1) in the main text, which involves an intermolecular SO effect, but they provide orientation on upper bounds, in the sense that we may assume that any energy scale related to intermolecular SO interactions should be smaller than any effects arising from intramolecular SO. This has been the rationale for our assumption in the main text to consider energy scales in the order of $\sim 10^{-4}$ meV for typical separations ~4 Å, which is much smaller than the typical energy scales we have shown so far. While this does not yield precise estimates for $\lambda_{eff}$, it shows that there is no need to assume unrealistically large values of the associated intermolecular SO energy scale. We finally mention that we are focusing here on a simple helical molecule. The results can be sensitively changed if we consider e.g. helical peptides, which have a more complex electrostatics as the pure-carbon based helicene, e.g. they possess large permanent dipole moments.

These DFT calculations clearly indicate that chiral discrimination is absent if dispersion forces are not explicitly included. However, this leaves unanswered the question as to whether there is a separate contribution arising from the CISS effect. The general question of separating exchange and spin-orbit contributions in a many-body description is an intricate one because these terms are not independent due to symmetry constraints on the wave function and the Hamiltonian itself. These considerations are especially important for our understanding of spin polarization phenomena in chiral systems, and justify the use of model calculations, like the one considered in the main text, which can shed light on this important topic.

### 3. Unsöld approximation

A commonly, though very rough, approximation to the London dispersion interactions is the so-called Unsöld approximation.[1] For the sake of completeness, we will show it here for the spin-spin interaction term. Going back to Eq. (6) in the main text, we have:

$$K_{\alpha\beta}^{BA} = -\lambda_{eff}^2 \varepsilon_{\alpha\kappa\gamma} \varepsilon_{\beta\kappa'\gamma'} T_{\gamma\rho} T_{\gamma'\rho'} \sum_{n,m\neq 0} \frac{\omega_{n0}\omega_{m0}}{\omega_{n0}+\omega_{m0}} \mu_\kappa^{0n}(A) \mu_\rho^{0m}(B) \mu_{\kappa'}^{m0}(B) \mu_{\rho'}^{n0}(A), \qquad (S10)$$

the standard Unsöld approximation is carried out as:

$$\frac{\omega_{n0}\omega_{m0}}{\omega_{n0}+\omega_{m0}} = \frac{U_A U_B}{U_A + U_B}(1+\Delta_{nm}).$$

If the correction term $\Delta_{nm}$ is small, then one can neglect it and one gets:

$$K_{\alpha\beta}^{BA} = -\lambda_{eff}^2 \frac{U_A U_B}{U_A + U_B} \varepsilon_{\alpha\kappa\gamma}\varepsilon_{\beta\kappa'\gamma'} T_{\gamma\rho} T_{\gamma'\rho'} \sum_{n\neq 0} \mu_\kappa^{0n}(A)\mu_{\rho'}^{n0}(A) \sum_{m\neq 0} \mu_\rho^{0m}(B)\mu_{\kappa'}^{m0}(B), \tag{S11}$$

The sums over states can be further simplified:

$$\sum_{n\neq 0} \mu_\kappa^{0n}(A)\mu_{\rho'}^{n0}(A) = \sum_{n\neq 0} \omega_{m0} \langle 0|\mu_\kappa(A)|n\rangle\langle n|\mu_{\rho'}(A)|0\rangle = \langle 0|\mu_\kappa(A)|(1-|0\rangle\langle 0|)\mu_{\rho'}(A)|0\rangle \tag{S12}$$

$$= \langle 0|\mu_\kappa(A)\mu_{\rho'}(A)|0\rangle - \langle 0|\mu_\kappa(A)|0\rangle\langle 0|\mu_{\rho'}(A)|0\rangle$$

$$= \langle 0|\mu_\kappa(A)\mu_{\rho'}(A)|0\rangle = \delta\mu_{\kappa\rho'}(A)$$

A similar expression holds for the other molecule $B$. In the last row we have further assumed, to be in line with the main text, that the molecules have no permanent dipole moment in their ground state. We thus see that at the level of the Unsöld approximation,

$$K_{\alpha\beta}^{BA} = -\lambda_{eff}^2 \frac{U_A U_B}{U_A + U_B} \varepsilon_{\alpha\kappa\gamma}\varepsilon_{\beta\kappa'\gamma'} T_{\gamma\rho} T_{\gamma'\rho'} \delta\mu_{\kappa\rho'}(A)\delta\mu_{\rho\kappa'}(B), \tag{S13}$$

the spin-spin interaction term $J_{\alpha\beta} = K_{\alpha\beta}^{BA} + K_{\alpha\beta}^{AB}$ is closely related to the quadratic fluctuations of the dipole operators in the ground state of the system. For a given orientation, e.g. $\vec{R} = (R,0,0)$, Eq. (S11) can be further simplified to (with $\vec{t} = (-2,1,1)$)

$$K_{\alpha\beta}^{BA} = -\frac{1}{R^6} C_{\alpha\beta}^{BA}. \tag{S14}$$

with $C_{\alpha\beta}^{BA} = \lambda_{eff}^2 U_A U_B /(U_A + U_B)\varepsilon_{\alpha\kappa\rho}\varepsilon_{\beta\kappa'\rho'} t_\rho t_{\rho'} \delta\mu_{\kappa\rho'}(A)\delta\mu_{\rho\kappa'}(B)$.